\documentclass[11pt]{article}
\usepackage{amsfonts}
\usepackage{amssymb}
\usepackage{latexsym}
\usepackage{graphicx}
\usepackage[english]{babel}
\begin{document}
\input epsf

\def\be{\begin{equation}}
\def\ee{\end{equation}}
\def\bea{\begin{eqnarray}}
\def\eea{\end{eqnarray}}
\def\p{\partial}
\def\r{\rightarrow}
\def\h{{1\over 2}}
\def\b{

\bigskip

}
\def\hb{Hawking believer:\quad}
\def\u{|\uparrow\rangle}
\def\d{|\downarrow\rangle}
\def\sq{{1\over \sqrt{2}}}
\def\z{|0\rangle}
\def\o{|1\rangle}
\def\sqi{{1\over \sqrt{2}}}

\begin{flushright}
\end{flushright}
\vspace{20mm}
\begin{center}
{\LARGE  The information paradox: A pedagogical introduction}
\\
\vspace{18mm}
{\bf  Samir D. Mathur\footnote{mathur@mps.ohio-state.edu} }\\

\vspace{8mm}
Department of Physics,\\ The Ohio State University,\\ Columbus,
OH 43210, USA\\
\vspace{4mm}
\end{center}
\vspace{10mm}
\thispagestyle{empty}
\begin{abstract}

The black hole information paradox is a very poorly understood problem. It is often believed that Hawking's argument is not precisely formulated, and a more careful accounting of naturally occurring quantum corrections will allow the radiation process to become unitary. We show that such is not the case, by proving that small corrections to the leading order Hawking computation cannot remove the entanglement between the radiation and the hole. We  formulate Hawking's argument as a `theorem': assuming `traditional' physics at the horizon and usual assumptions of locality we {\it will} be forced into mixed states or remnants. We also argue that one cannot explain away the problem by   invoking AdS/CFT duality. We conclude with recent results on the quantum physics of black holes which show the the interior of black holes have a `fuzzball' structure. This nontrivial structure of microstates resolves the  information paradox, and gives a qualitative picture of how classical intuition can break down in black hole physics.

\end{abstract}
\vskip 1.0 true in

\newpage
\setcounter{page}{1}

\section{The black hole information paradox paradox}

Consider the following conversation between two students:

\bigskip

First student: \quad `Suppose I am falling into a black hole. By the equivalence principle, I will notice nothing special
at the horizon.  Hawking has argued that radiation from such a horizon has a thermal spectrum, which means that the radiation can carry no information and information will be lost when the black hole evaporates.'

\bigskip

Second student: \quad `I know that this is called the information paradox, but I think there are many weaknesses in Hawking's argument. The radiation he computes starts as transplankian modes, so it is wrong to treat those modes like free field modes, as Hawking did. In any case there is no information paradox in string theory because we know that black holes in AdS at least are dual to a field theory, and the field theory is manifestly unitary.'

\bigskip

I will argue in these notes that each phrase in the above conversation is either incorrect or is irrelevant to the information problem. Some forty years after Hawking became famous for finding this problem, one of the most paradoxical things is that the black hole information paradox is so well known and yet so poorly understood. A principal goal of these lectures will therefore be to understand exactly what the paradox {\it is}. We will see that there is a very precise statement of the contradiction found by Hawking, and that bypassing the paradox needs a basic change in our understanding of how quantum effects operate in gravity.  

\section{The information paradox}

We develop the ideas underlying the paradox in a number of steps, starting with very basic principles. It is essential to understand these principles, because even though we use them automatically in our everyday use of physics, it is these same principles which are being challenged by the information paradox.

\subsection{Existence of a `solar system physics' limit}

General relativity has brought several complications to physics. In particular, space and time can `stretch', and we are then forced to extend quantum theory to curved spacetime. In our solar system we have space-time curvature, but we  believe that we can do laboratory experiments without worrying about details of quantum gravity.  Why is that? The vacuum of a free quantum field in flat space has fourier modes of all wavelengths going down to zero. When quantum gravity is taken into account, what will be the behavior of modes with wavelengths shorter than planck length? We do not fully understand the answer to such questions, and if it were the case that the full physics of quantum gravity were needed for describing {\it any} process, then we would not be able to do much physics. 

The reason that we do not worry about quantum gravity in everyday experiments is that we believe there is an appropriate limit where the effects of quantum gravity become small, and a local, well defined, approximate evolution equation becomes possible. The existence of this limit is so crucial to physics, that I use  a specific name for it in these notes:  `solar system physics'. The term  signifies that we can do normal physics when spacetime curvatures are of the order found in our solar system. 

\b

{\bf Definition D1: Solar system limit: } \quad {\it There must exist a set of `niceness conditions' N containing a small parameter $\epsilon$ such that when $\epsilon$ is made arbitrarily small then physics can be described to arbitrarily high accuracy by a known,  local,  evolution equation. That is, under conditions N we can specify the quantum state on an initial spacelike slice, and then a Hamiltonian evolution operator gives the state on later slices. Furthermore, the influence of the state in one region on the evolution in another region must go to zero as the distance between these regions goes to infinity (locality).}

\b

We have not yet specified these niceness conditions N, but the fact that there must exist such a `solar system limit' underlies all of our physical thinking. Hawking's theorem starts with a natural set of niceness conditions N, and proves that  requiring locality with these conditions would lead to an `unacceptable' physical evolution. One must then either agree to this `unacceptable evolution', or find a way to add new conditions to the set N, in such a way that these conditions still allow us to define a  `solar system limit' incorporating some idea of locality.

\subsection{`Niceness' Conditions' N for local evolution}\label{n}

Here we make a list of `niceness conditions' that are traditionally assumed, whether explicitly or implicitly:

\b

(N1) Our quantum state is defined on a spacelike slice. The intrinsic curvature ${}^{(3)} R$ of this slice should be much smaller than planck scale everywhere: ${}^{(3)} R\ll {1\over l_p^2}$. 

\b

(N2) The spacelike slice sits in an 4-dimensional spacetime. Let us require that the slice be nicely embedded in the full spacetime; i.e., the {\it extrinsic} curvature of the slice $K$ is small everywhere: $K\ll {1\over l_p^2}$. 

\b

(N3) The 4-curvature curvature of the full spacetime in the neighbourhood of the slice should be small everywhere ${}^{(4)} R\ll {1\over l_p^2}$

\b

(N4) We should require that all matter on the slice be `good'. Thus any quanta  on the slice should have wavelength much longer than planck length ($\lambda\gg l_p$), and the  energy density $U$ and momentum density $P$ should be small everywhere compared to planck density: $U\ll l_p^{-4}, ~~~P\ll l_p^{-4}$. 
Let us add here that we will let all matter satisfy the usual energy conditions (say, the dominant energy condition). 

\b

(N5) We will evolve the state on the initial slice to a later slice; all slices encountered should be `good' as above. Further, the lapse and shift vectors needed to specify the evolution should change smoothly with position: $
{dN^i\over ds}\ll {1\over l_p}, ~~~~{dN\over ds}\ll {1\over l_p} $

\b

\b

Note that not all these conditions may be   independent, but we have tried to be as generous as possible in making sure that we are in a domain of semiclassical physics. 

\b

\b

\subsection{The quantum process of interest}

\b

While we have tried to make sure that the spacetime is such that  quantum gravity can be ignored, we have to do a little more:  we have to be more precise about what physical process we are interested in looking at. The reason is simple. There is always {\it some} effect of quantum gravity, so if we insist on making a very precise measurement of something or look at some very abstruse physical effect, we might pick up effects of quantum gravity even in gently curved spacetime. So we will now pick the physical process of interest.

\bigskip

\begin{figure}[htbp]
\begin{center}
\includegraphics[scale=.15]{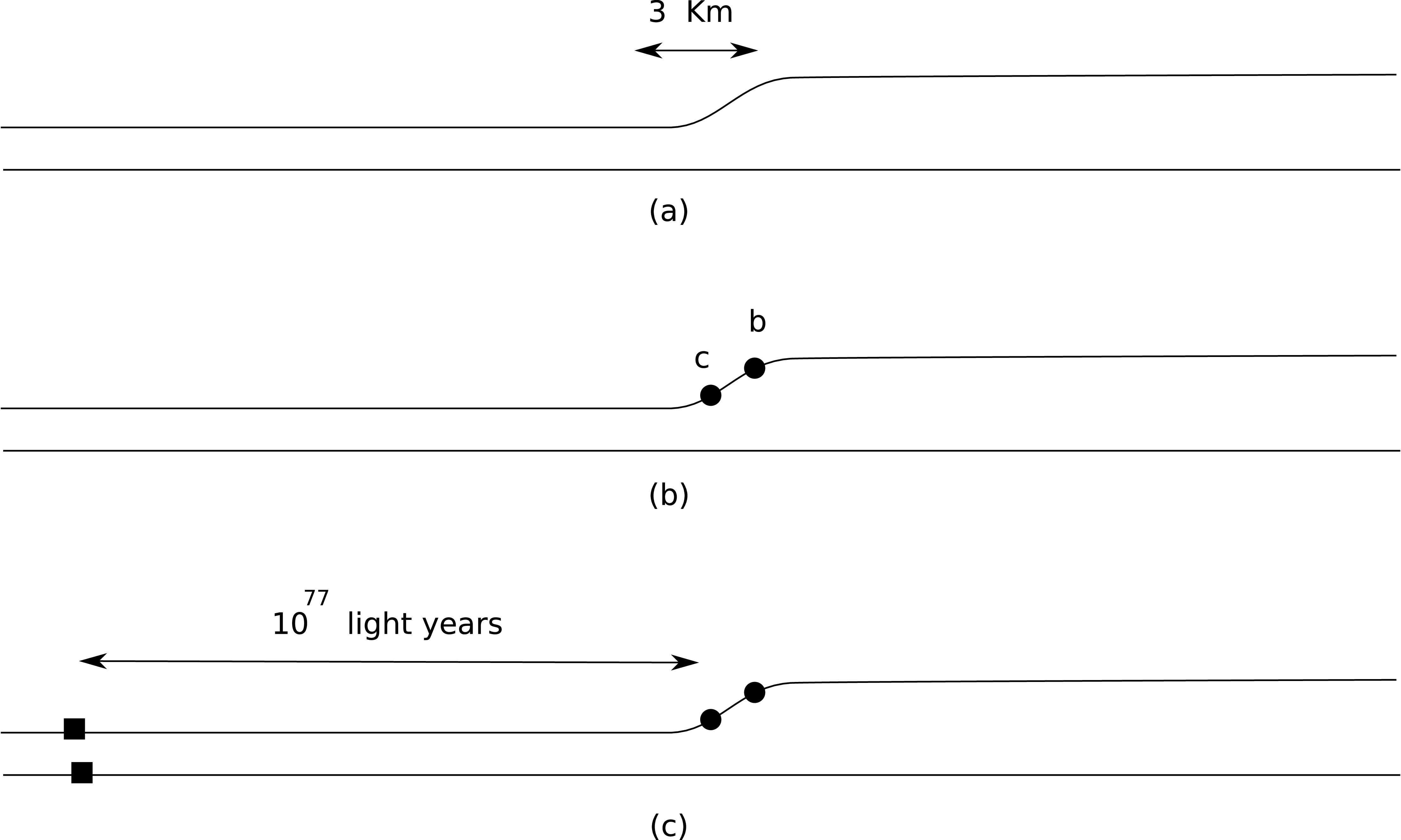}
\caption{(a) Spacelike slices in an evolution; the intrinsic geometry of the slice distorts in the region between the right and left sides (b) Particle pairs are created in the region of distortion (c) There is matter far away from the region of distortion; locality would say that the state of this matter is only weakly correlated with the state of the pair.}
\label{fone}
\end{center}
\end{figure}

{\bf 1.  A set of nice slices:}  Start with the vacuum state on the lower slice in fig.\ref{fone}(a). Consider the evolution to the upper slice shown in the figure.  The later slice is evolved forward in the right hand region, less so in the left hand region. This is of course allowed: in general relativity we have `many-fingered time' in the language of Wheeler, so we can evolve in any way that we like. The slices satisfy all the `niceness' conditions N. 

\b

{\bf 2. Pair creation:} The evolution of the geometry will lead to particle creation in the region where the geometry of the slice is being deformed; this happens because  the vacuum state on one slice will not in general be the natural vacuum on a later slice.  Let the geometry in the deformation region 
 be characterized by the length and time scale $L$. Then the particle pairs created have wavelengths $\lambda\sim L$
and the number of such created pairs is $n\sim 1$. For concreteness we will take 
$L\sim 3 ~{\rm Km}$
which is the curvature length scale at the horizon of a solar mass black hole.

The particle pair is depicted by $c, b$ in figure fig.\ref{fone}(b).
The state of the created pairs is of the form
 \be
 |\Psi\rangle_{\rm pair}=Ce^{\gamma{\hat c} ^\dagger {\hat b}^\dagger}|0\rangle_c|0\rangle_b
 \label{actualstate}
 \ee
 where $\gamma$ is a number of order unity. The detailed form of this state can be found in \cite{hawking,giddings}. But the essence of this entanglement can be obtained by assuming the following simple form for the state
\be
|\Psi\rangle_{\rm pair}=\sq \z_c\z_b+\sq\o_c\o_b
\label{pairs}
\ee
where in fig.\ref{fone}(b)  the quantum of the right is called $b$ and the quantum on the left is called $c$.

\b

{\bf 3. Matter on the slice:} There is some matter in a state $|\psi\rangle_M$ on the spacelike slice, but the crucial point is that this matter is very far away, a distance
$L'\gg L$
from the place where the pair creation is taking place. For concreteness, we will take
$L'\sim 10^{77} ~~{\rm light~years}$ (fig.\ref{fone}(c)).
(This is the length scale that we will encounter in the Hawking evaporation problem of the solar mass hole.)

\b

{\bf 4. Locality:} If we now assume  `locality' on the spacelike slices then the complete state on the spacelike slice would be
\be
|\Psi\rangle\approx |\psi\rangle_M\otimes\Big( \sq \z_c\z_b+\sq\o_c\o_b\Big)
\label{qtwo}
\ee
Even though the matter $|\psi\rangle_M$ is far away from the place where the pairs are being created, there will always be {\it some} effect of $|\psi\rangle_M$ on the state of the create pairs. This is why we have written an $\approx$ sign in (\ref{qtwo}). But it is crucial to Hawking's argument that we make this more precise.

Let the state of matter $|\psi\rangle_M$ consist of a single spin which can be up or down. Let us take
$
|\psi\rangle_M=\Big(\sq|\uparrow\rangle_M+\sq|\downarrow\rangle_M\Big )
\label{tw}
$.
Then if there was no effect of the matter state $|\psi\rangle_M$ on the state of the pairs, the state on the slice would be
\be
|\Psi\rangle\approx \Big(\sq|\uparrow\rangle_M+\sq|\downarrow\rangle_M\Big )\otimes\Big( \sq \z_c\z_b+\sq\o_c\o_b\Big)
\label{qtwop}
\ee
It is crucial to understand that locality allows small departures from the state (\ref{qtwop}), for example
\be
|\Psi\rangle=  \Big (\sq|\uparrow\rangle_M+\sq|\downarrow\rangle_M\Big )\otimes\Big [({1\over \sqrt{2}}+\epsilon)\z_c\z_b+({1\over \sqrt{2}}-\epsilon)\o_c\o_b\Big ], ~~~~\epsilon\ll 1
\label{qthree}
\ee
but not a completely different state like
\be
|\Psi\rangle=\Big(\sq\u_M\z_{c}+\sq\d_M\o_{c}\Big )\otimes \Big (\sq\z_{b}+\sq \o_{b}\Big )
\label{qfour}
\ee

\b

{\bf 5. Quantifying locality:} Let us now quantify the sense in which (\ref{qthree}) is close to (\ref{qtwop}) but (\ref{qfour}) is not. For the state (\ref{qtwop}) write the density matrix describing the quantum $b$, while tracing over the degrees of freedom in $M$ and $c$. Note that $b$ is entangled with $c$, but not with $M$. One finds the entanglement entropy
\be
S_{ent}=-tr [\rho\ln \rho]=\ln 2
\ee
Doing the same computation for the state (\ref{qthree}) gives
\be
S_{ent}=-tr [\rho\ln \rho]=\ln 2-\epsilon^2(6-2\ln 2)\approx \ln 2
\ee
But with the state (\ref{qfour}) we get
\be
S_{ent}=0
\ee

  Now we can make a precise statement about the consequence of assuming locality. Consider the limit
  \be
{L\over l_p}\gg1, ~~~{L'\over l_p}\gg 1, ~~~{L'\over L}\gg 1
\label{limit1}
\ee
The first two inequalities say that all length scales are much longer than planck length, and the last  says that the matter $M$ is `far away' from the place where the pairs are being created. Note that both the length and time scales involved in the pair creation process are $\sim L$. 
The above limits are amply satisfied by the scales we took in fig.\ref{fone}.
\b
Now we can express the notion of locality in the following  way
\b
{\it If we assume that the niceness conditions N give `solar system physics', then  in the limit} (\ref{limit1}) {\it we will get}
\be
{S_{ent}\over \ln 2}-1\ll 1
\ee

\section{The `traditional black hole'}\label{trhole}

\b

The above discussion was very general, but let us now apply it to the black hole with metric
\be
ds^2=-(1-{2M\over r}) dt^2+(1-{2M\over r})^{-1} dr^2 + r^2 d\Omega_2^2
\label{ten}
\ee
This metric has a horizon at $r=2M$. The curvature in the vicinity of the horizon is low: ${}^4R\sim {1\over M^2}$. We will call this the `traditional black hole geometry'. The essential feature of the traditional geometry that will be of relevance to us is the fact that  there is no `information' about the hole in the vicinity of the horizon. We call this an `information-free horizon' and make this precise with the following definition:

\b

{\bf Definition D2:} \quad {\it A point on the horizon will be called `information-free' if around this point we can find a neighborhood which is the `vacuum' in the following sense: the evolution of field modes with wavelengths $l_p\ll \lambda\lesssim M$ is given by the semiclassical evolution of quantum fields on `empty' curved space upto terms that vanish as ${m_p/ M}\r 0$.}

\b

{\it A black hole with an infomation-free horizon will be called a `traditional black hole'.}

\b

 Note that in any curved space there is no unique definition of particles, but if the curvature radius is $R$ then for wavemodes with wavelength $\lambda \lesssim R$ we {\it can} get a definition of particles in which we can say what `empty space' is. That is, one definition of particle will disagree with another but only in that the two definitions will differ by $\sim 1$  quanta with $\lambda \sim R$.  At the black hole horizon we have the same situation, and one might expect to see order unity quanta with $\lambda\sim M$ in any natural definition of particles; Hawking radiation quanta that are being created will also fall in this category.  Thus `empty' in the above definition means that we do not have for instance modes with $\lambda \lesssim M/10$ populated by $\sim 1$ quanta each.
 
 Finally, note that the metric (\ref{ten}) gives a time-independent black hole geometry. Hawking radiation makes the black hole slowly evaporate, but since the evaporation process is slow, we can describe the traditional black hole by a  metric like (\ref{ten}) at any given point of the evaporation process except near the very end when the black hole becomes planck sized.
 
\b

\b

\subsection{Slicing the traditional  black hole geometry}

\b

The traditional black hole has a spacelike singularity inside the horizon. If this singularity intersects the spacelike slices of our evolution, then the niceness conditions N would not hold everywhere on the slice, and one would not be able to make  Hawking's argument. Thus it is crucial that we can make a set of spacelike slices that describe a spacetime region ${\cal S}$, satisfy all the niceness conditions N, and yet capture all the physics of Hawking radiation. We now describe this very important construction of slices.

\subsubsection{Making one spacelike slice}

We make a spacelike slice for the black hole (\ref{ten}) as follows (fig.\ref{ftwo}):

\b

(a) For $r>4M$ we let the slice be $
t=t_1=constant
$

\b

(b) Inside $r<2M$ the spacelike slices are $r=constant$ rather than $t=constant$. We let the slice be
$
r=r_1, ~~{M/ 2}<r_1<{3M/ 2}
$,
so that this part of the slice is not near the horizon  $r=2M$ and not near the singularity  $r=0$.

\b

(c) We join these parts of the slice with a smooth `connector' segment ${\cal C}$. It is easy to check that such a connector can be made so that the slice satisfies all the niceness conditions N which we gave above.

\b

\begin{figure}[htbp]
\begin{center}
\includegraphics[scale=.25]{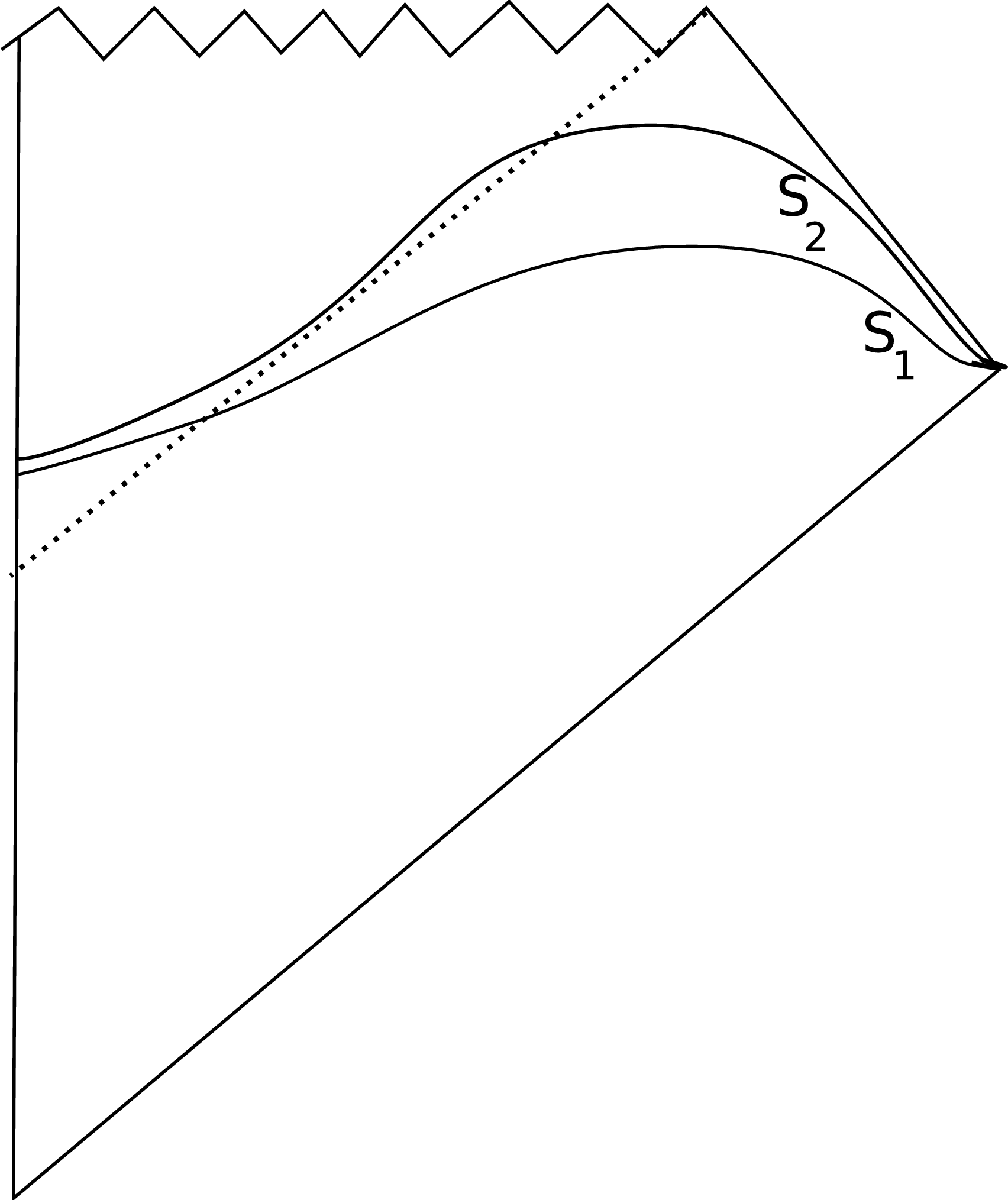}
\caption{{The Penrose diagram of a black hole formed by collapse of the `infalling matter'. The spacelike slices satisfy all the niceness conditions N.}}
\label{fthree}
\end{center}
\end{figure}

(d) The geometry (\ref{ten}) gives the time independent black hole, but we will be interested in black holes made by starting with flat space and having a shell of mass $M$ collapse towards the origin $r=0$. The Penrose diagram for this hole is shown in fig.\ref{fthree}. With such a hole made by collapse we can follow the $r=r_1$ part of the slice down to early times before the hole was formed, and then smoothly extend it to $r=0$ (there is no singularity at $r=0$ at these early times).

\subsection{Evolution to later slices}

\begin{figure}[htbp]
\begin{center}
\includegraphics[scale=.18]{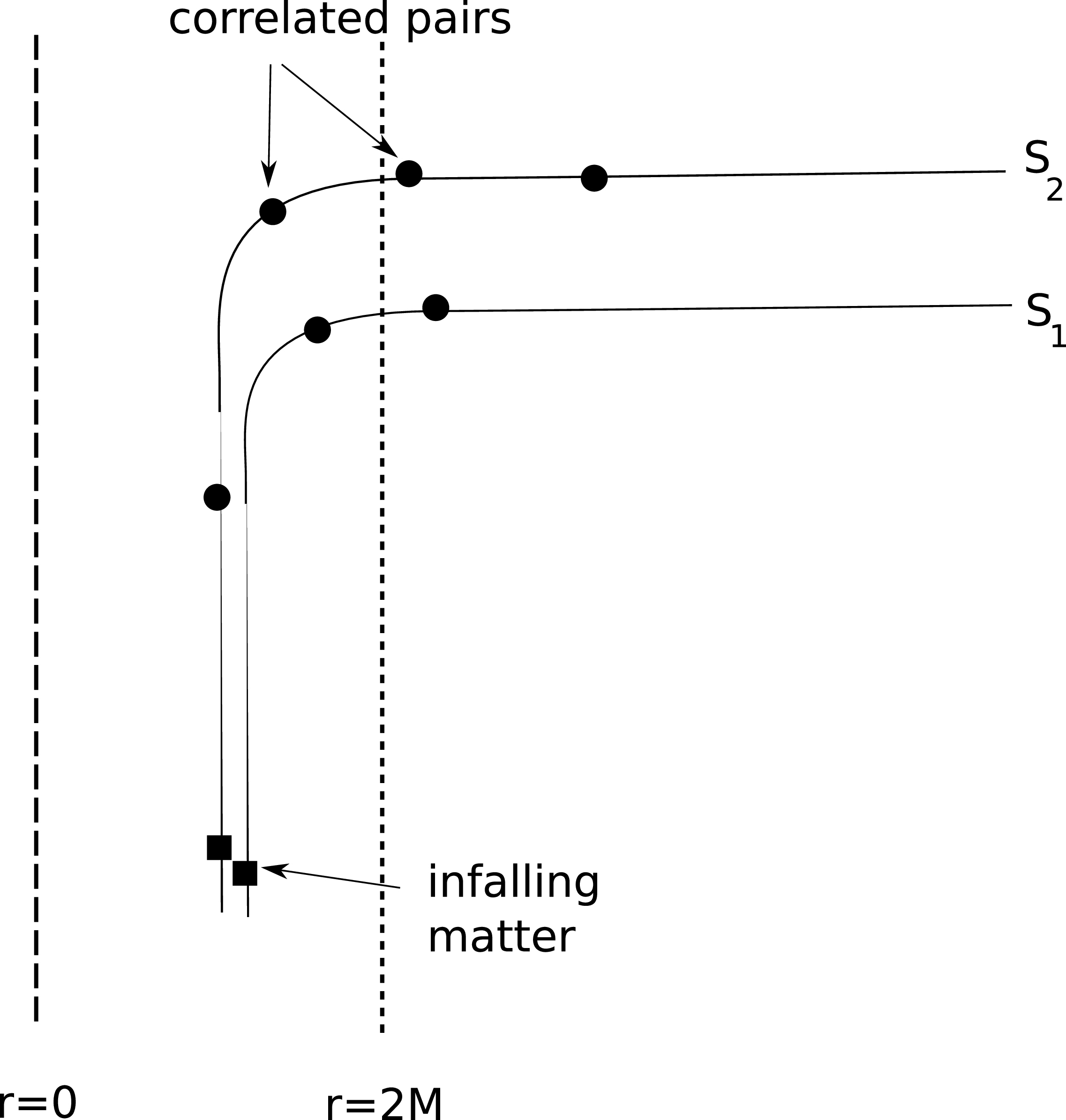}
\caption{{A schematic set of coordinates for the Schwarzschild hole. Spacelike slices are $t=const$ outside the horizon and $r=const$ inside. Infalling matter is very far from the place where pairs are created ($\sim 10^{77}$ light years) when we measure distances along the slice. Curvature length scale is $\sim 3 ~km$ all over the region of evolution covered by the slices $S_i$.}}
\label{ftwo}
\end{center}
\end{figure}

This makes one complete spacelike slice, which we call as ${ S}_1$ in fig.\ref{ftwo}. Let us now make a   `later' spacelike slice ${ S}_2$.

\b

(a)  At  $r>4M$  we take $t=t_1+\Delta$. 

\b

(b) The $r=const$ part will be
$r=r_1+\delta$
where $\delta_1\ll M$. Note that the timelike direction for this part of the geometry is in the decreasing $r$ direction, so this change $\delta_1$ is indeed correct for evolution  of this part of the slice.  We let $\delta_1$ be small, and will later take the limit where $\delta_1\r 0$ for convenience. 

\b

(c) We again join the parts (a),(b) by  a smooth `connector' segment. In the limit $\delta_1\r 0$ we can take the geometry of the connector segment ${\cal C}$ to be the same for all slices. Note the very important fact (which can be seen from fig.\ref{ftwo}) that the $r=const$ part of the later slice ${S}_2$ is {\it longer} than the $r=const$ part of ${ S}_1$. This extra part of the slice is needed because the connector segment has to join the $r=const$ part to the $t=const$ part, and the $t=const$ part has been evolved forwards on the later slice.  

\b

(d) At early times we again bring the $r=const$ part smoothly down to $r=0$, at a place where there is no singularity. 

\subsection{The changes between the slices}\label{slicing}

Let us now see the nature of the evolution from slice ${ S}_1$ to slice ${S}_2$. Let us choose lapse and shift vectors on the spacetime as follows. We take the slice ${ S}_1$ and pick a point $x^i$ on it. Now move along the timelike normal till we reach a point on ${ S}_2$. Let  this point on ${ S}_2$ have the same spatial coordinates  $x^i$. Thus we have set the shift vector to be $N^i=0$. With this choice we can describe the evolution as follows:

\b

(a)  In the $t=constant$ part of the slice we have no change in intrinsic geometry. This part of the slice just advances forward in time with a lapse function $N=(1-{2M\over r})^\h$.

\b

(b)  Let us work in the limit $\delta_1\r 0$. The $r=const$ part of ${ S}_1$ moves over to ${S}_2$ with no change in intrinsic geometry. The early time part which joins this segment to $r=0$ also remains unchanged. 

\b

(c) The connector segment ${\cal C}$ of ${ S}_1$ has to `stretch' during this evolution, since the corresponding points on ${ S}_2$ will have to cover both  the connector ${\cal C}$ of ${ S}_2$ and the extra part of the $r=const$ segment possessed by ${ S}_2$. 

\b

Thus we see that the `stretching' happens only in the region near the connector segment. This region has space and time dimensions of order $M$, which is $3~km$ for the solar mass black hole that we have taken as our example. 

Now consider a succession of slices ${ S}_n$ of this type, evolving from ${ S}_n$ to ${ S}_{n+1}$ in exactly the same way that we evolved from ${ S}_1$ to ${ S}_2$. Each evolution from ${ S}_n$ to ${ S}_{n+1}$ can be described as follows:

\b

{\it Divide ${ S}_n$ into a left part, a right part, and a middle part (which is the connector region). In the evolution to  ${ S}_{n+1}$, the left and right parts stay unchanged but are pushed apart, and the middle part is stretched to a longer length. The length of the middle part is $\sim M$, the proper time between the slices in this middle region is $\sim M$, and the stretching is by a factor $(1+\alpha)$, with $\alpha\sim 1$.}

\b

One can now check the following important fact: 

\b

{\bf Nice slicing of the traditional hole: } \quad {\it  The slicing constructed above satisfies all the `niceness conditions' $N$ listed in section \ref{n} above.}

\b

In fact these slices look just like the ones shown in fig.\ref{fone}.   The curvature is low in all the regions covered by the evolution. Note that while the metric (\ref{ten}) looks time independent, this is only an illusion because the Schwarzschild coordinates break down at the horizon; any slicing that covers both the outside and the inside of the hole will necessarily be time dependent.
This time-dependence is the underlying reason for particle creation in the black hole geometry.

\subsection{A comment on the `stretching' in this slicing}

The slicing of the black hole geometry is very interesting. All the `stretching' between successive slices happens in a given place, so that the  fourier modes of fields at this location keep getting stretched to larger wavelengths, and particles will keep being produced (until the hole becomes very small). We cannot have such a set of slices in ordinary Minkowski space. If we try to make slices like those in fig.\ref{ftwo} in Minkowski space, then after some point in the evolution the later slices will not be spacelike everywhere: the `stretching' part will become null and then timelike. But it is the basic feature of the black hole that the space and time directions interchange roles inside the horizon, and we get spacelike slices having a stretching like that of fig.\ref{ftwo} throughout the region of interest.

Looking at such a slicing one might wonder if there should be a problem with stretching a given region of spacetime `too much'. If we could make such a notion concrete, then we could add a new condition to the niceness conditions N, prohibiting `too much stretching'. Then  the slicing used in the derivation of Hawking's theorem would cease to satisfy the niceness conditions, and the information paradox would be bypassed. But if we add such a condition, then we will have to face its consequences. Our Universe starts with a small marble sized ball, and its spatial sections stretch without bound. If we propose that `too much stretching' destroys the `solar system limit', then we have to consider that that this limit will be violated at some point in Cosmological evolution. (In \cite{stretch} such a condition was proposed, and its consequence for Cosmological evolution was explored in \cite{cosmo}.)  Thus we see that it is not easy to add new conditions to the set N without suffering serious modifications to physics in other situations. 

A second possibility is to argue that while the slicing of the spacetime region itself satisfies the niceness conditions N, there is a singularity in the future of the region covered by the slices, and this could affect the evolution on the slices and generate nonlocal effects. Again, if one wishes to make this argument, one has to make a precise statement of when the niceness conditions will be violated by a singularity in the future. Suppose one does manage to add such a condition to the niceness set N. Then what will happen if our Universe approaches a Big Crunch? There is a singularity in the future, so will we get nonlocal effects all across our spacelike slices? Also, by time reversal symmetry, if we had a singularity in the past then should it not have similar effects? In that case, just after the Big Bang we should get nonlocal effects across the spacelike slices of our Universe, something that we do not build into our normal understanding of Cosmology. So we see again that it is not easy to add new conditions to the set N. In the discussion below we will  just take the standard set of niceness conditions given in section \ref{n} in exploring the information paradox.

\section{The Hawking radiation process: leading order}\label{haw}

\b

The evolution of slices in the black hole geometry will lead to the creation of particle pairs. The members of these pairs that float out to infinity are called Hawking radiation. In this section we look at the nature of the wavefunction of these created pairs. The pairs will form a state which is {\it entangled} in a very specific way, and this fact will form the heart of Hawking's theorem. It is crucial that the state of these pairs is a state unlike any that is created when a normal hot body radiates photons. We will see that the essential difference arises from the fact that in the black hole case the particle pairs are the result of `stretching' of a region of the spacelike slice; thus these pairs are `pulled out of the vacuum'. In normal hot bodies the radiation is emitted from the constituents making up the hot body. This is the essential difference between a hot body and the black hole, so we will be returning to this issue repeatedly.

In this section we will see the state of the created pairs `to leading order'. If this were the actual state of the pairs, it is easy to see that we cannot escape Hawking's conclusion. But of course there will always be small corrections to any leading order computation. A common misconception is that these small corrections can remove Hawking's problem and make the black hole radiation no different from radiation from a hot body. {\it This is absolutely incorrect}. Thus  in proving  Hawking's theorem we will explore the space of small corrections to the leading order state and show that {\it unless we make an order unity modification to the leading order result, Hawking's problem will persist}. Well, why not assume then that there are order unity corrections to the lading order state? The crucial point of Hawking's theorem is that in that case {\it there will be breakdown of the solar system limit even if we are given the niceness conditions N}. 

This then is the outline of the Hawking result. It would clearly be very difficult to write down all sources of small corrections to the leading order result. The proof of the theorem does away with the necessity of analyzing these corrections, by simply  mapping the  evolution to one in a domain of `solar system physics', so that any effect that generates corrections large enough to make the hole radiate like a normal body will be forced to lead to a violation of the solar system limit. 

Clearly, we need to get a good understanding of evolution in the black hole spacetime. To make the essence of the proof clearer, we start in this section with the leading order Hawking state, analyze its essential properties, and postpose the investigation of corrections to the next section.

\b
\begin{figure}[htbp]
\begin{center}
\includegraphics[scale=.25]{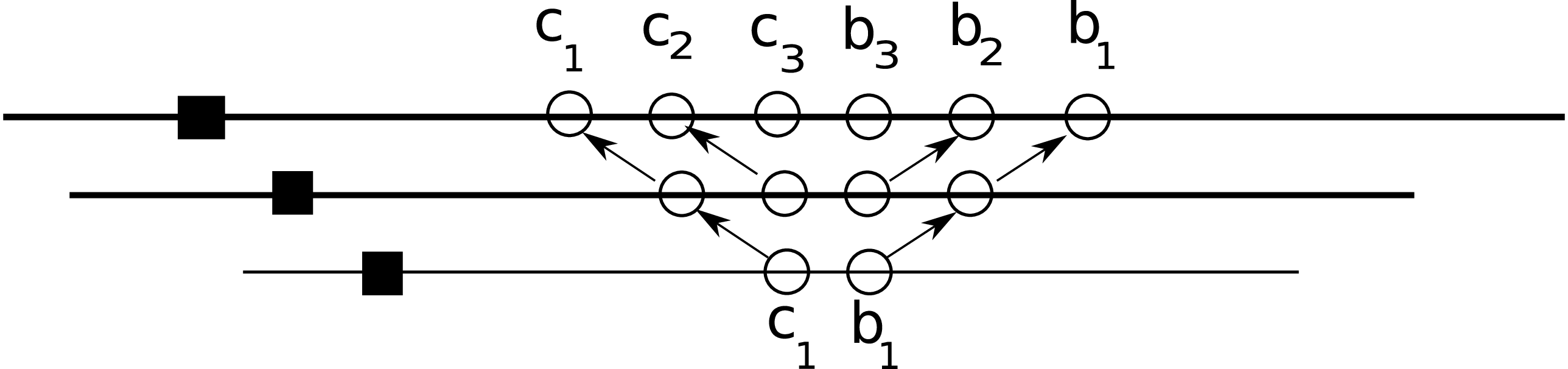}
\caption{{The creation of Hawking pairs. The new quanta $c_{n+1}, b_{n+1}$ are not created by interaction with either the matter $|\psi\rangle_M$ (represented by the black square) or with the earlier created pairs. Rather the creation is by a Schwinger process which moves $|\psi\rangle_M$ further away from the place of pair creation, and also moves the earlier created $c, b$ quanta away from the place of pair creation. The new pairs are created in a state which to leading order is entangled between the new $b,c$ quanta but not entangled with anything else. Small corrections to this leading order state does not change this entanglement significantly, so the entanglement keeps growing all through the radiation process, unlike the case of radiation from normal hot bodies.}}
\label{ffour}
\end{center}
\end{figure}

(1) Consider an initial spacelike slice. The shell that collapses to make the hole is represented by a matter state $|\psi\rangle_M$. 

\b

(2) Let us evolve to the next spacelike slice. The `middle part' of the spacelike slice stretches, while the left and right parts remain unchanged. This is shown in fig.\ref{ffour}. 
The stretching creates  correlated pairs (labelled $b_1, c_1$) and the state on the complete slice is like (\ref{qtwo})
\be
|\Psi\rangle\approx |\psi\rangle_M\otimes\Big( \sq \z_{c_1}\z_{b_1}+\sq\o_{c_1}\o_{b_1}\Big)
\label{qtwoq}
\ee
If we compute the entanglement of $b_1$ with $M, c_1$ we obtain 
\be
S_{entanglement}=\ln 2
\ee

\b

(3) Now consider the next slice of spacetime in fig.\ref{ffour}. The following happens during this step of the evolution:
\b

(i) The matter state $|\psi\rangle_M$ stays almost the same because there is no evolution in this part of the slice
\b
(ii) The change in the geometry happens only  in the region where the `connector' part joins the $r=const$ part; in this region there is a `stretching' of the spacelike slice.  There are two consequences of this stretching. The first is that the pairs $b_1, c_1$ created earlier move away from each other and from the region of stretching. The second is that the stretching creates a new set of pairs $b_2, c_2$ in the region of stretching. Again, the full state of the quantum field can be found in \cite{hawking, giddings}, but for our present purposes we can write the state at the end of this step as
\bea
|\Psi\rangle\approx |\psi\rangle_M&\otimes&\Big( \sq \z_{c_1}\z_{b_1}+\sq\o_{c_1}\o_{b_1}\Big)\cr
&\otimes&\Big( \sq \z_{c_2}\z_{b_2}+\sq\o_{c_2}\o_{b_2}\Big)
\label{qtwoq2}
\eea
We compute the entanglement of the set $\{b_1, b_2\}$ with $\{M, c_1, c_2\}$. We get
\be
S_{entanglement}=2\ln 2
\label{ent2}
\ee

\b

(4) After $N$ such steps we will have the state
\bea
|\Psi\rangle\approx |\psi\rangle_M&\otimes&\Big( \sq \z_{c_1}\z_{b_1}+\sq\o_{c_1}\o_{b_1}\Big)\cr
&\otimes&\Big( \sq \z_{c_2}\z_{b_2}+\sq\o_{c_2}\o_{b_2}\Big)\cr
&\dots&\cr
&\otimes&\Big( \sq \z_{c_N}\z_{b_N}+\sq\o_{c_N}\o_{b_N}\Big)
\label{qtwoq3}
\eea
The space $\{ b_i\}$ is entangled with the remainder $M, \{ c_i\}$ with
\be
S_{entanglement}=N\ln 2
\label{ent}
\ee

\b

(5) As the quanta $\{ b_i\}$ collect at infinity, the mass of the hole decreases. The slicing does not satisfy the niceness conditions N after the point when
$
M_{hole}\sim m_{pl}
$
because ${}^4R\ll l_p^{-2}$ is no longer true. We will therefore stop evolving our spacelike slices when this point is reached.

\b

The emitted radiation quanta $\{b_i\}$ have an entanglement (\ref{ent}) with $M, \{ c_i\}$. We define:

\b

{\bf Definition D3:} {\it We will say that our gravity theory contains {\it remnants} if there exists a set of  objects with mass and size less than given bounds 
\be
m<m_{remnant}, ~~~l<l_{remnant}
\ee
but allowing an arbitrarily high entanglement with systems far away from the object.}

\b

It is easy to see that an object can be entangled with $S_{entanglement}=n\ln 2$ with another system only if the number of possible states of the object is $\gtrsim n$. Thus remnants have unbounded degeneracy while having energy and size within given bounds. This is {\it not} the expected behavior of quantum systems in normal physics. It has been argued that existence of remnants gives rise to loop divergences from the infinite number of possible objects circulating in the loop. 

We now see that our slicing of the traditional black hole geometry has forced us to the choose between the following possibilities:

\b

{\bf (1) Mixed states}:\quad  The black hole evaporates away completely. The quanta $\{ b_i\}$ have entanglement entropy
$
S_{ent}\sim N\ln 2\ne 0
\label{enentropy}
$.
But since there is nothing left that they are entangled with, the final state is not described by any quantum wavefunction. The final state can only be described  by a density matrix.

\b

{\bf (2) Remnants}\quad  The evolution stops when $M_{hole}\sim m_{remnant}$. Given the entanglement (\ref{ent}), we find the theory contains remnants.
\b

Possibility (1) leads to a loss of unitarity of quantum mechanics, since a pure initial state evolved to a mixed final state. Possibility (2) is not a violation of quantum mechanics as such, but still differs from expected physics, since normally we expect only a finite number of states if both the energy and spatial extent are bounded.   The Hawking argument cannot say which of possibilities (1) and (2) will occur since the niceness conditions $N$ are violated near the endpoint of evolution. Thus we will lump both these possibilities into one which we call `mixed states/remnants'.

\b

As mentioned at the start of this section, the above discussion is not really the full Hawking theorem yet because we have not considered possible small corrections that will in general exist to the state (\ref{qtwoq3}). Thus what we have seen so far is a first outline of the Hawking argument. To recapitulate this outline, what we have seen is that at each stage of the evolution the entanglement entropy of the $b_i$ in this leading order state {\it increases} by $\ln 2$. The evolution is very unique to the black hole because the radiation is created by the stretching of the `middle part' of each spacelike slice. It is the peculiar nature of the black hole geometry which creates such persistent stretching of spacelike slices. When normal hot bodies radiate, the radiation quanta are {\it not} created by stretching of spacelike slices. Thus for normal hot bodies the radiation quanta depend on the nature of the atomic state at the surface of the body. By contrast, in  black hole evolution  we see that the matter making the hole ($|\psi\rangle_M$ in the above discussion) stays far away from the place where the Hawking pairs are being created. In fact with each successive stage of stretching, this matter is moved further {\it away} from the place where the next pair would be produced. 

How far away is this matter for the creation of the typical Hawking pair for a solar mass black hole? After each stage of stretching, the matter moves a distance of order $ M\sim 3~km$ away from the place where the pairs are being created. The number of radiation quanta is $(M/m_p)^2$. Thus after about half the evolution, the distance of the matter (measured along the spacelike slice) from the place where the pairs are being created is of order
\be
L\sim  M ({M\over m_p})^2\sim 10^{77} ~ {\rm light ~years}
\ee
This shows the sharp contrast between the case of normal hot bodies and the case of the black hole. For normal hot bodies, the distance $L$ between the matter in the body and the place where the radiation is created would be {\it zero} since the radiation leaves from the atoms in the body.

One might think that even though the matter $|\psi\rangle_M$ is very far away from where the pairs are being created, the pairs which have been created recently are close to the new pair being created, and this may help to generate correlations. Again, one finds that this does not happen. For one thing, the earlier created $b,c$ quanta also move away from the pair creation region at each step. Thus the typical $b$ or $c$ quantum is also a distance of order $L\sim 10^{77}$ light years from the place where the new pairs are being created. Of course the pairs that have been {\it recently} created are at a  distance $\sim 3 ~km$ from the newly created pair. But the nature of the pair creation process is such that this nearness does not help. The new pair is created by the stretching of a {\it new} fourier mode, and the earlier created pair is simply pushed away in this process. To see this fact more explicitly, one has to write down the actual wavemodes and the wavefunctional describing their overall state. The reader who wishes to get a full understanding of this point should take the full wavefunctional given in \cite{giddings} and use it to understand why the simplified state (\ref{qtwoq3}) gives the same essential physics of correlations as the full wavefunctional.

\section{Stability of the Hawking state}

We saw above the essence of the Hawking argument: repeated stretching at the horizon leads to repeated creation of pairs in a  particular entangled state (\ref{qtwoq3}). This forces us to mixed states/remnants.

There will of course be small corrections to the leading order state (\ref{qtwoq3}). These can come from small interactions between earlier created pairs and the newly created pair. The matter shell which made the hole $|\psi\rangle_M$ is very far away on the spacelike slice, but the entire black hole is a gravitational solution with radius $M$, and there can be instanton effects that lead to small effects of $|\psi\rangle_M$ on the pair being created. Such instanton effects would be exponential small
\be
{\cal A}_{instanton}\sim e^{-S_{instanton}}, ~~~S_{instanton}\approx GM^2
\ee
where we have used the action of the standard instanton found in black hole physics: the Euclidean solution
\be
ds^2=(1-{2M\over r}) d\tau^2+(1-{2M\over r})^{-1} dr^2 + r^2 d\Omega_2^2
\label{tenq}
\ee
with identification $\tau=\tau +{8\pi G M}$. 

It is logical to ask if such small corrections can perhaps invalidate Hawking's argument, and remove all entanglement between the quanta $b_i$ and the $(M, c_i)$ quanta in the hole. If this happens, there would be no paradox, since the hole containing $(M,c)$ can vanish, and we will be left with a pure state of the $b_i$ quanta, presumably carrying all the information of the initial matter $|\psi\rangle_M$. 

It is a very important fact that such small corrections do {\it not} change the conclusion reached in the Hawking argument. We will call this fact the `stability of entanglement of the Hawking state'. In this section we will first define the meaning of small corrections, then prove a set of three lemmas that will  lead to the proof of the  stability theorem. In the next section we will use this theorem to state and prove Hawking's theorem on mixed states/remnants.

\subsection{Deformations of the leading order state}\label{deformation}

Let the state of the modes in the box at time step $t_n$ be written as  $|\Psi_{M, c}, \psi_{b}(t_n)\rangle$. Here $\Psi_{M,c}$ denotes the state of the matter shell that fell in to make the black hole, and also all the $c$ quanta that have been created at earlier steps in the evolution. $\psi_b$ denotes the set of all $b$ quanta that have been created in all earlier steps. This state is entangled between the $(M,c)$ and $(b)$ parts; it is not a product state. We assume nothing about its detailed structure. 

 In the leading order evolution we would have  at time step $t_{n+1}$:
\be
|\Psi_{M,c}, \psi_b(t_n)\rangle\r |\Psi_{M,c}, \psi_{b}(t_n)\rangle~\Big [\sqi |0\rangle_{c_{n+1}}|0\rangle_{b_{n+1}}+\sqi |1\rangle_{c_{n+1}}|1\rangle_{b_{n+1}}\Big ]
\label{leading}
\ee
where the term in box brackets denotes the state of the newly created pair. 

In this section we will write down the most general deformation of this leading order state, and analyze the entanglement entropy that results. Thus in a sense the computations here are at the heart of the Hawking theorem, since they will prove that it is not possible to avoid mixed states/remnants by the small corrections which will always be present to a leading order result like (\ref{leading}).

\b

(1) We have seen that the stretching of spacetime creates a new region on the spacelike slice, and moves the regions to the left and right of this region apart, without distortion. The full state of the created pairs is given in \cite{giddings}, but we will continue to use our simplified model of this state: we allow only occupation number of a mode to only be $0$ or $1$, and we discretize the evolution into steps where pairs get created in one mode at each step. These simplifications allow a presentation that is possible to follow easily. (If the reader is worried that some generality has been lost then he should start with the full state in \cite{giddings} and repeat the steps below.)

We assume that  state in the newly created region is spanned by 2 vectors, for which we can take the basis
\bea
S^{(1)}&=&\sqi  |0\rangle_{c_{n+1}}|0\rangle_{b_{n+1}}+\sqi  |1\rangle_{c_{n+1}}|1\rangle_{b_{n+1}}\cr
S^{(2)}&=&\sqi  |0\rangle_{c_{n+1}}|0\rangle_{b_{n+1}}-\sqi  |1\rangle_{c_{n+1}}|1\rangle_{b_{n+1}}\cr
&&
\label{set}
\eea
We could have included other vectors to enlarge this space, but the nature of the argument that follows would not change, so we leave it to the reader to change the equations below to reflect a larger choice of possibilities if he wishes.  

\b

(2) The state at time $t_n$ is $|\Psi_{M,c}, \psi_b(t_n)\rangle$. This has entanglement between the the $b_i$ and the state inside the hole, which is composed of the matter making the shell ($M$) and the infalling members of pairs $c_i$ that have been created till now. We can choose a basis of orthonormal states $\psi_n$ for the $(M,c)$ quanta inside the hole, and an orthonormal basis $\chi_n$ for the quanta $b_i$ outside the hole, such that 
\be
|\Psi_{M,c}, \psi_b(t_n)\rangle=\sum_{m,n} C_{mn} \psi_m\chi_n
\ee
It is  convenient to make unitary transformations on the $\psi_i, \chi_j$  so that we get
\be
|\Psi_{M,c}, \psi_b(t_n)\rangle=\sum_{i} C_{i} \psi_i\chi_i
\label{stateone}
\ee
We compute the reduced density matrix describing the $b_i$ quanta outside the hole
\be
\rho_{ij}=|C_i|^2 \delta_{ij}
\ee
The entanglement at time $t_n$ is then
\be
S_{ent}(t_n)=-\sum_i |C_i|^2 \ln |C_i|^2
\label{entone}
\ee

\b

(3) We now consider the evolution to the next timestep $t_{n+1}$. The only constraint we put on the evolution is that the $b_i$ which have been created in all earlier steps be not affected by this step of the evolution. This is because these $b_i$ quanta have left the vicinity of the hole, and can be collected outside; in any case they can no longer be influenced by the hole without invoking some magical `action at a distance' that would violate any locality condition that we assume. We can write  these quanta $b_i$ in terms of creation operators for outgoing modes, and in this basis the state does not evolve further (the quanta keep moving out to larger $r$, but this is along a line of fixed $t-r$). 

(If the reader wishes to argue that the most recent $b_i$ quanta are still not far enough from the hole and can be influenced, he should try to build the argument in detail using a larger space than (\ref{set}) which includes several pairs instead of just one.)

Thus the most general evolution to timestep $t_{n+1}$ is given by 
\be
\chi_i\r \chi_i
\ee
\be
\psi_i\r \psi_i^{(1)}S^{(1)}+\psi_i^{(2)}S^{(2)}
\ee
where the state $\psi_i$ of the $(M, c_i)$ evolves to a tensor product of states $\psi_i^{(i)}$ describing $(M,c_i)$ and  the $S^{(i)}$ which described the newly created pair.  Note that unitarity of evolution gives
\be
|| \psi_i^{(1)}||^2+|| \psi_i^{(2)}||^2=1
\ee
In the leading order evolution we had
\be
\psi_i^{(1)}=\psi_i, ~~\psi_i^{(2)}=0
\ee
and we will use this fact below to define what we mean by `small' corrections and `order unity' corrections.

Putting all this together,  the state (\ref{stateone}) evolves at $t_{n+1}$ to
\be
|\Psi_{M,c}, \psi_b(t_{n+1})\rangle=\sum_{i} C_{i} [\psi_i^{(1)}S^{(1)}+\psi_i^{(2)}S^{(2)}]~ \chi_i
\label{statetwo}
\ee

\b

(4) We  wish to compute the entanglement entropy in the state (\ref{statetwo}), between the $b$ quanta that are outside the hole and the $(M,c)$ quanta that are inside the hole. The $b$ quanta outside include the $b_i$ from all steps upto $t_{n}$ as well as the quantum $b_{n+1}$ created in this last step of evolution. We now wish compute $S_{ent}(t_{n+1})$ in terms of $S_{ent}(t_n)$ given by (\ref{entone}) and the states $\psi_i^{(\alpha)}$ in (\ref{statetwo}). 

\b

We now define explicitly what we mean by  a `small' change in the evolution of a new pair. Write the state (\ref{statetwo}) as
\be
|\Psi_{M,c}, \psi_b(t_{n+1})\rangle=S^{(1)}\Big [ \sum_{i} C_{i} \psi_i^{(1)} \chi_i\Big ] + S^{(2)}\Big [ \sum_{i} C_{i} \psi_i^{(2)} \chi_i\Big ] \equiv S^{(1)} \Lambda^{(1)}+ S^{(2)} \Lambda^{(2)}
\label{state}
\ee
Here we have defined the states
\be
\Lambda^{(1)}= \sum_{i} C_{i} \psi_i^{(1)} \chi_i, ~~\Lambda^{(2)}= \sum_{i} C_{i} \psi_i^{(2)} \chi_i
\ee
Since $S^{(1)}, S^{(2)}$ are orthonormal, normalization of $|\Psi_{M,c}, \psi_b(t_{n+1})\rangle$ implies that
\be
||\Lambda^{(1)}||^2+||\Lambda^{(2)}||^2=1
\ee

\b

{\bf Definition D2:} \quad 
We will call corrections small if  
\be
|| \Lambda^{(2)}||<\epsilon, ~~~\epsilon\ll 1
\label{cond1}
\ee
If there is no such bound, then we will say that the corrections are `order unity'.

\b

The physical significance of this definition is simple to understand. When put the  in the context of the black hole problem, we have seen that the creation of the new pair  occurs in a spacetime satisfying the niceness conditions N. We will see in the proof of the Hawking theorem below  that if the traditional black hole structure is assumed at the horizon, then the effect of quanta already present on the spacelike slice is supposed to be small under these conditions N. This means that we can reliably find the state $S^{(1)}$ when the pair is created, and the probability to find the orthogonal state $S^{(2)}$ has to be much less than unity. Thus we get (\ref{cond1}) as the definition of small corrections. Note that this condition still allows that for special states $\psi_i$ of the hole we can have a large amplitude for $S^{(2)}$; this is still consistent with the requirement of small corrections if   $C_i$ be sufficiently small. Thus (\ref{cond1}) says that corrections are small if for probable states of the hole we have close to unity probability for generating the new pair in the state $S^{(1)}$.

\subsection{Entropy bounds}

The entropy at step $t_n$ of the evolution is given by (\ref{entone}). Let us call this value $S_0$. Our goal is to show that at step $t_{n+1}$ the entropy of entanglement of the $b$ quanta with the quanta in the hole $(M, c)$ {\it increases}
if we satisfy the requirement of `small corrections'  defined above. It may seem obvious that if corrections are small then the increase in entropy will be close to the value $\ln 2$ that we find in the leading order evolution. The reason that this may not be completely clear though is that there is a very large number of $c$ quanta that the newly created pair $(c_{n+1}, b_{n+1})$ can entangle with, and one might think that a very delicate entanglement with this large number of quanta may allow the overall entanglement entropy to go down while still allowing the state of $(c_{n+1}, b_{n+1})$ to be close to the one demanded by the niceness conditions. Here we will prove three lemmas leading to the following theorem:\footnote{I thank Patrick Hayden for suggesting the use of entropy inequalities in deriving this result.}  if we are given the bound (\ref{cond1}), then at each stage of the evolution the entanglement entropy will increase by  at least $\ln 2-2\epsilon$. As can be imagined, this theorem will then be crucial to establishing the Hawking argument that the niceness conditions N force mixed states/remnants.

  \b

Let us recall some standard notation. Suppose we have a system with subsystems $A, B, C$. (We assume that the overall system is in a pure state, though this is not needed for the inequalities below.) Then $S(A)\equiv -tr \rho_A\ln \rho_A$  where $\rho_A$ is the density matrix describing subsystem $A$. $S(A)$ thus gives the entropy of entanglement of subsystem $A$ with the rest of the system. Similarly $S(A+B)$ is the density matrix of the union of subsystems $A$ and $B$; it gives the entanglement of $A+B$ with $C$.

Our black hole system has the following subsystems:

\b

(i) The radiation quanta $\{ b_1, \dots b_n \}$ emitted in all steps upto and including step $t_n$. We call this set  $\{ b\}$. As noted above, we always assume that quanta emitted at earlier steps do not participate in the dynamics of pair creation at the next timestep.
\b
(ii) The black hole interior contains at timestep $t_n$ the shell state $|\psi\rangle_M$ and the quanta $\{ c_1\dots c_n\}$. We lump these together into the symbol $(M, \{ c\})$. The pair created at timestep $t_{n+1}$ can interact weakly with $(M, \{c\})$ creating entanglements which were not present  in the leading order Hawking state;
these are the effects that we are trying to consider now.
\b
(iii) The pair $p$ that will be created at the timestep $t_{n+1}$. We write this as
$p\equiv (c_{n+1}, b_{n+1})$. 

\b

Now consider the entropies of these subsystems. 
 Let the entropy (\ref{entone}) be called $S_0$. Thus at time step $t_n$ we have
$
S_{\{ b\}}=S_0
$.
Our assumption that the earlier emitted quanta $\{ b\}$ cannot be influenced any further implies that even at the timestep $t_{n+1}$ we continue to have
\be
S_{\{ b\}}=S_0
\label{s0}
\ee

\b

Our final goal will be to show that $S(\{b\}, b_{n+1})>S_0 -2\epsilon$, so that we establish that despite small corrections, the entanglement entropy {\it increases} {\it} at the timestep $t_{n+1}$. First we prove
\b
 
{\bf Lemma L1:}\quad If (\ref{cond1}) holds, then the entanglement of the pair $(c_{n+1}, b_{n+1})$ with the rest of the system is bounded as
\be
S(c_{n+1}, b_{n+1})\equiv -tr \rho_{(c_{n+1}, b_{n+1})}\ln \rho_{(c_{n+1}, b_{n+1})} < \epsilon
\ee
\b
{\it Proof:}\quad The density matrix for the system $(c_{n+1}, b_{n+1})$ is
\be
\rho_{(c_{n+1}, b_{n+1})}=\pmatrix { \langle \Lambda^{(1)}|\Lambda^{(1)}\rangle & \langle \Lambda^{(1)}|\Lambda^{(2)}\rangle\cr \langle \Lambda^{(2)}|\Lambda^{(1)}\rangle&\langle \Lambda^{(2)}|\Lambda^{(2)}\rangle\cr }
\label{density}
\ee
From (\ref{cond1}) we are given that
\be
|| \Lambda^{(2)}||^2=\langle \Lambda^{(2)}|\Lambda^{(2)}\rangle\equiv \epsilon_1^2<\epsilon^2
\label{ineq}
\ee
Then by the Schwartz inequality
\be
|\langle \Lambda^{(1)}|\Lambda^{(2)}\rangle|\equiv \epsilon_2<\epsilon
\ee
Now note that if we have a density matrix
\be
\rho=\h I +\vec \alpha\cdot \vec\sigma
\ee
then we can make a unitary transformation to bring it to the form
\be
\rho=\h I + |\vec \alpha| \sigma_3
\ee
The entropy of this density matrix is then seen to be
\be
S=-tr\rho\ln \rho=\ln 2 -\h (1+2|\vec \alpha| )\ln (1+ 2 |\vec \alpha| )-\h (1-2|\vec \alpha| )\ln (1- 2 |\vec \alpha| )
\ee
Applying this relation to (\ref{density}), with (\ref{ineq}), we find
\be
S(c_{n+1}, b_{n+1})=(\epsilon_1^2-\epsilon_2^2)\ln {e\over (\epsilon_1^2-\epsilon_2^2)}+O(\epsilon^3)<\epsilon
\label{lemma1}
\ee
for $\epsilon\ll 1$. \quad $\square$

\b
We write the result of the above lemma as 
\be
S_p<\epsilon
\label{sp}
\ee
where $p=(b_{n+1}, c_{n+1})$ denotes the pair created at timestep $t_{n+1}$. This result shows that the entire pair $p$ is weakly entangled with the remainder of the system.  Next we prove

\b

{\bf Lemma L2:} \quad 
\be
S({\{b\}+p})\ge S_0-\epsilon
\label{lemma2}
\ee

\b

{\it Proof:}\quad We use the subadditivity property relating the entropy of two systems $A, B$
\be
S(A+B)\ge |S(A)-S(B)|
\ee
Let $A=\{ b\}$ and $B=p$. Then (\ref{lemma2}) follows immediately from (\ref{s0}),(\ref{sp}). \quad $\square$

\b 

{\bf Lemma L3:}\quad
\be
S_{c_{n+1}}>\ln 2-\epsilon
\label{sc}
\ee

\b
{\it Proof:}\quad To prove this look at the state (\ref{state}) and write it in the form
\be
|\Psi_{M,c}, \psi_b(t_{n+1})\rangle=\Big[|0\rangle_{c_{n+1}}|0\rangle_{b_{n+1}}\sqi(\Lambda^{(1)}+\Lambda^{(2)})\Big ]+ \Big [|1\rangle_{c_{n+1}} |1\rangle_{b_{n+1}}\sqi(\Lambda^{(1)}-\Lambda^{(2)})\Big ]
\ee
The density matrix describing $c_{n+1}$ is
\be
\rho_{c_{n+1}}=\pmatrix{ \h\langle (\Lambda^{(1)}+\Lambda^{(2)})|(\Lambda^{(1)}+\Lambda^{(2)})\rangle & 
0\cr 0& \h\langle (\Lambda^{(1)}-\Lambda^{(2)})|(\Lambda^{(1)}-\Lambda^{(2)})\rangle\cr}
\ee
Using $\langle\Lambda^{(2)}|\Lambda^{(2)}\rangle= \epsilon_2^2$, $\langle\Lambda^{(1)}|\Lambda^{(1)}\rangle=1-\epsilon_2^2$,  we get
\be
\rho_{c_{n+1}}=\h I+\pmatrix{ Re\langle\Lambda^{(1)}|\Lambda^{(2)}\rangle&0\cr 0&-Re\langle\Lambda^{(1)}|\Lambda^{(2)}\rangle\cr}+O(\epsilon^2)
\ee
and
\be
S(c_{n+1})=\ln 2 -2[Re(\langle\Lambda^{(1)}|\Lambda^{(2)}\rangle)]^2\ge \ln 2 -2\epsilon^2 +O(\epsilon^3)>\ln 2-\epsilon
\ee
for $\epsilon\ll 1$. 

\b

We now use the above lemmas to prove the stability of entanglement of the leading order Hawking state:
\b
 
{\bf Theorem 1:}\quad 
Suppose at time step $t_n$ the quanta $\{ b_1\dots b_n\}$ have been emitted, and their total entanglement entropy with the hole is $S_0$. Suppose the next pair emitted at timestep $t_{n+1}$ departs from the leading order Hawking state $\sqi(|0\rangle_{c_{n+1}}|0\rangle _{b_{n+1}}+
|0\rangle_{c_{n+1}}|0\rangle _{b_{n+1}})$ by an amount less than  $\epsilon\ll 1$; this condition being defined precisely in   (\ref{cond1}). Then after this time step the entropy of the emitted quanta $\{ b_1, \dots b_{n+1}\}$ will satisfy
\be
S(\{b\}+b_{n+1})> S_0+\ln 2-2\epsilon
\ee
Thus the entanglement entropy of the emitted quanta necessarily {\it increases} with each emission if the departures from the leading order Hawking state are small.

\b
{\it Proof:}\quad 
We use the strong subadditivity theorem \cite{lieb} which governs the entropies of three systems
 \be
 S(A+B)+S(B+C)\ge S(A)+S(C)
 \ee
 We set $A=\{ b \}, B= b_{n+1}, C=c_{n+1}$. This gives
 \be
 S(\{ b\}+b_{n+1})+S(p)\ge S(\{ b \})+S(c_{n+1})
 \ee
We have from (\ref{sp}) that $S(p)<\epsilon$. From (\ref{s0}) we have $S(\{ b \})=S_0$. From (\ref{sc}) we have $S(c_{n+1})>\ln 2-\epsilon$. This gives
\be
S(\{ b\}+b_{n+1})>S_0+\ln 2 -2\epsilon ~~~~~~~\square
\ee
\b

This is the result that we wished to establish. It proves that whenever we have `normal physics' at the horizon (i.e. the parameter giving departures from the leading order state is $\epsilon\ll 1$)
then the entanglement always {\it increases} by at least $\ln 2-2\epsilon$ after each stage of evolution. The entanglement thus cannot go down at any of these stages. Thus small corrections do not change the conclusion of Hawking's leading order result, and we will get mixed states/remnants when the traditional black hole evaporates, unless we can show that the corrections to evolution are really order unity and not order $\epsilon\ll 1$.

\b

Why have we taken so much trouble to establish this `stability of entropy increase'? 
A principal confusion about black holes arises from the following (erroneous) argument: 

\b

(a) When a piece of paper burns away the information is captured by the radiation, but it is very hard to read this information from the radiation because it is encoded in delicate correlations between radiation quanta. In fact Page \cite{page} has shown that almost no information about the paper can be read off unless we look at a subsystem including at least half the radiation quanta.

\b

(b) Because the information is encoded in delicate correlations, a similar situation can hold in the case of Hawking radiation. Hawking computed only the leading order state, and did not look at tiny corrections to this state. Even though these corrections may be only exponentially small (rather than order unity), they can be enough to solve the information paradox because only very delicate correlations were needed anyway.

\b

\begin{figure}[htbp]
\begin{center}
\includegraphics[scale=.20]{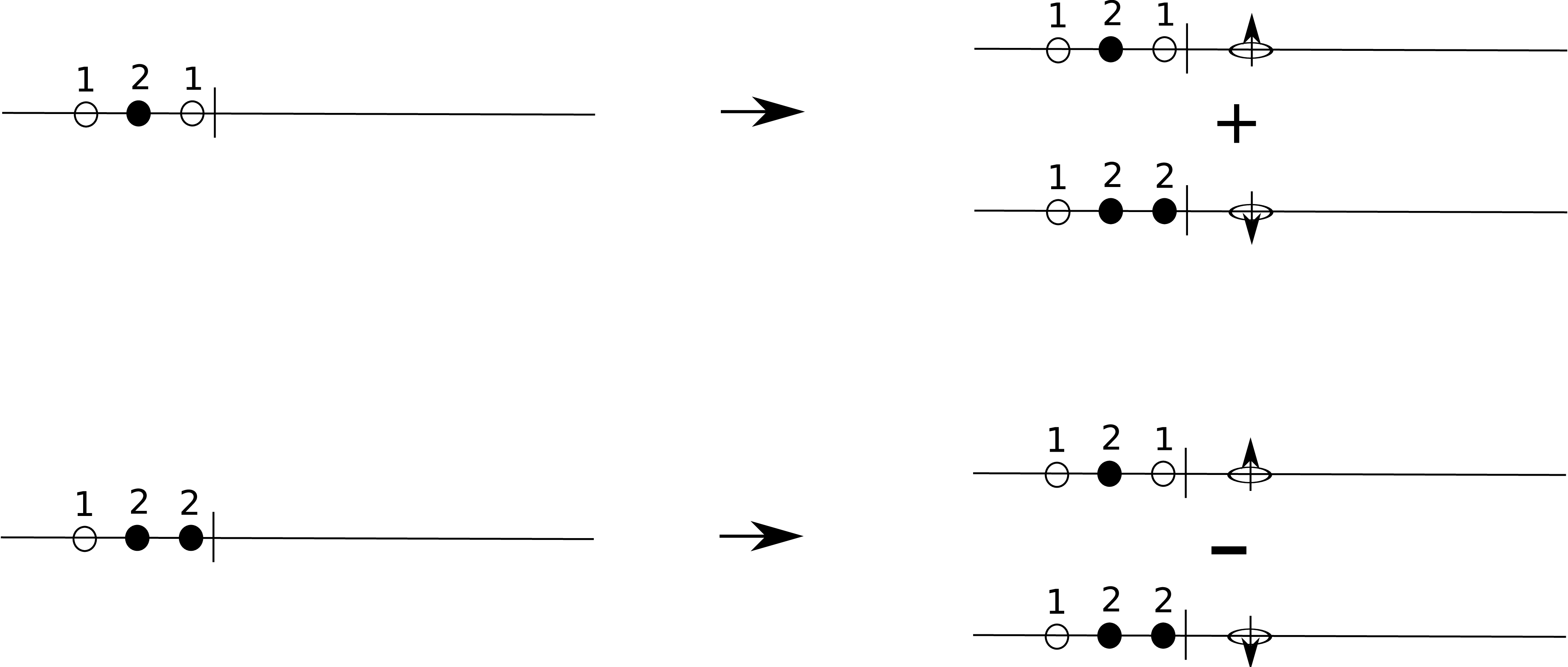}
\caption{{ A toy model of radiation from a normal hot body, showing the evolutions given in 
eqs.(\ref{first}),(\ref{second}). The left side of the vertical bar shows atoms in the body, the right side shows radiated photons with the arrow depicting their spin. If the atom near the boundary is in state $1$ (unfilled circle) then we get the linear combination of states on the right, and if the atom is in state $2$ (filled circle), then we get an orthogonal different linear combination.}}
\label{ffive}
\end{center}
\end{figure}

While (a) is correct, the followup step (b) is completely wrong, as we see by Theorem 1. Let us pinpoint the error more precisely. Consider a toy model of the burning paper. If the atom on the surface is in state $|1\rangle a$, then  let this evolve to an entangled state of atom plus radiated photon as follows
\be
|1\rangle_a~\r~\sqi |1\rangle_a|\uparrow\rangle_{ph}+\sqi |2\rangle_a|\downarrow\rangle_{ph}\equiv S^{(1)}
\label{first}
\ee
where $|2\rangle_a$ is another state of the atom orthogonal to $|1\rangle_a$, and the spin states of the photon are given with a subscript `$ph$'. If the atom was in state $|2\rangle_a$, let this evolve as
\be
|2\rangle_a~\r~\sqi |1\rangle_a|\uparrow\rangle_{ph}-\sqi |2\rangle_a|\downarrow\rangle_{ph}\equiv S^{(2)}
\label{second}
\ee
Note that $S^{(2)}$ is orthogonal to $S^{(1)}$ as it must be, since a unitary evolution will map orthogonal states $|1\rangle_a, |2\rangle_a$ to orthogonal states. In fig.\ref{ffive} we show the evolution of this system: the state produced at time step $t_{n+1}$ depends, {\it to leading order}, on the state at time step $t_n$. Thus in this toy model we get with equal probability the states
\be
S^{(1)}, ~~S^{(1)}; ~~~\langle S^{(1)}| S^{(2)}\rangle=0
\label{dimension}
\ee
By contrast, the black hole case was sketched in fig.\ref{ffour}. In each time  step 
we create, to leading order, the {\it same} state $S^{(1)}$. This is the crucial difference between the black hole case and the radiation from any hot body. The emission from the hot body must necessarily be an interaction leading to a vector space of states with dimension $d>1$ ; we have $d=2$ in the above toy example (\ref{dimension}). We get one state or the other from this vector space depending on the state of the atom near the surface. Of course interactions within the body can make this emission much more complicated, but this requirement $d>1$ is what allows the information to come out. By contrast, the black hole case has $d=1$ to leading order, so the state produced at the interaction point is the {\it same} regardless of the state of the black hole.  This leads to growing entanglement with each emission, and small corrections do not change the conclusion, as the above theorem shows.

Thus the essential mistake that one makes in step (b) of the above argument is in not specifying more carefully the term `delicate correlations'. There is nothing delicate about these correlations in the emission process from a  hot body: the emitted state changes radically from $S^{(1)}$ to an orthogonal state $S^{(2)}$ depending on the nature of the radiating atom. What {\it is} delicate is the difficulty of extracting the state of the burning paper from the state of the radiation, since the data of the state is shared among many radiation quanta. In the black hole the state of the created pair does {\it not} change radically from one emission to the next; the corrections away from the leading orders state $S^{(1)}$ are themselves delicate (i.e., very small), and this leads to a very different outcome from the hot body case. 

\section{The Hawking `Theorem'}

We now have all the tools needed to establish 
\b

{\bf Hawking's `Theorem':}\quad  {\it If we assume

\b

(i)  The Niceness conditions $N$ give local Hamiltonian evolution

\b

(ii) A traditional black hole (i.e. one with an information-free horizon) exists in the theory

\b

Then formation and evaporation of such a hole will lead to mixed states/remnants. }

\b
\b

{\it Proof:} \quad We proceed in the following steps:
\b

(1) Consider the metric (\ref{ten}) of the traditional black hole. By the observation noted in section  \ref{slicing} this black hole admits a slicing satisfying the niceness conditions N in the domain of interest. By assumption (i) of the theorem, this implies that we have `solar system physics' in the region around the horizon where particle pairs will be created.

\b

\begin{figure}[ht]
\includegraphics[scale=.20]{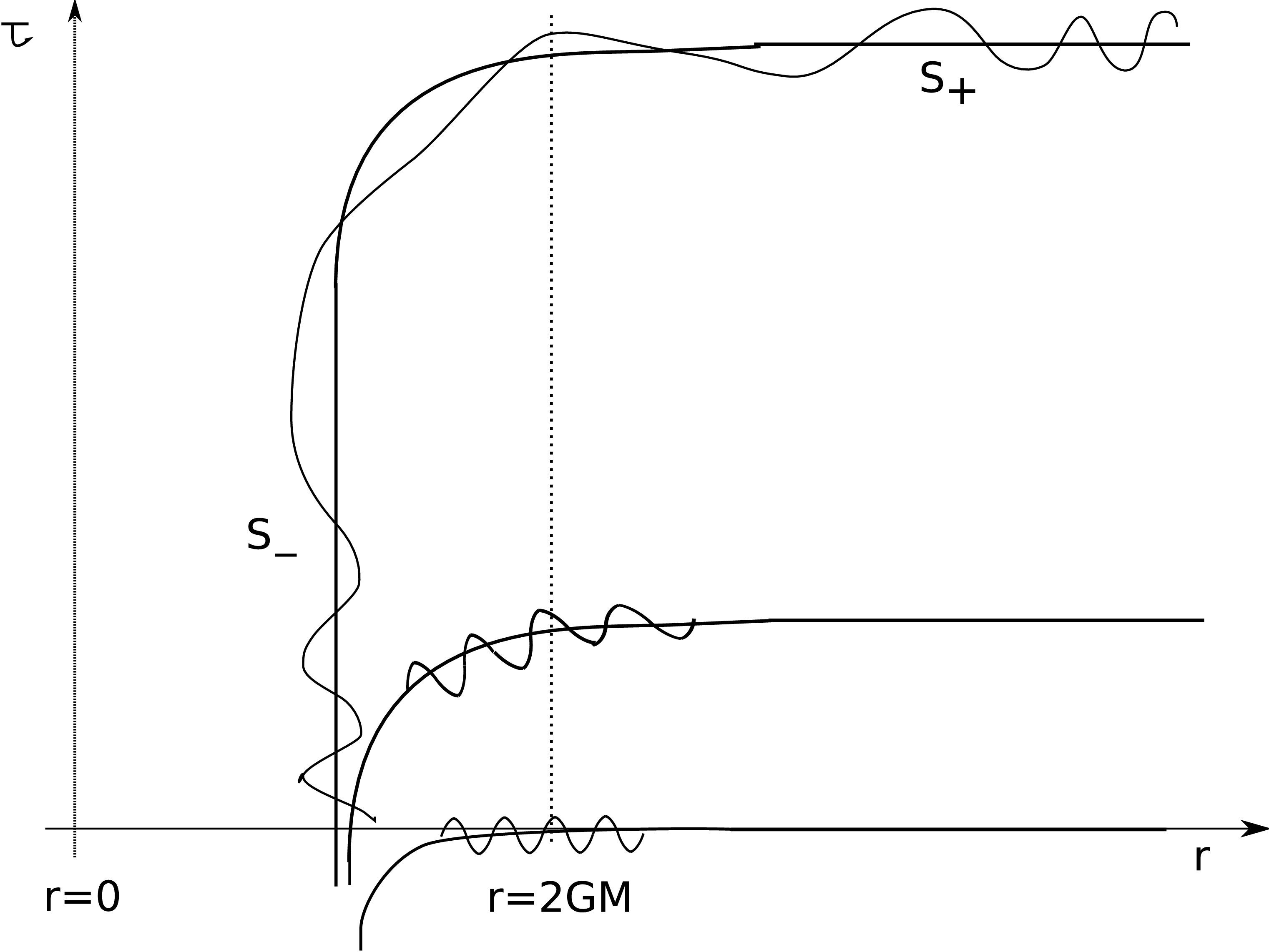}
%
%
\caption{A fourier mode on the initial spacelike surface is evolved  to later spacelike surfaces. In the initial part of the evolution the wavelength increases but there is no significant distortion of the general shape of the mode. At this stage the initial vacuum state is still a vacuum state. Further evolution leads to a distorted waveform, which results in particle creation. ($\tau$ is a schematic time coordinate; since this is not a Penrose diagram illustrating the actual spacetime structure of the geometry.)}
\label{matffourt}       
\end{figure}

\begin{figure}[ht]
\includegraphics[scale=.20]{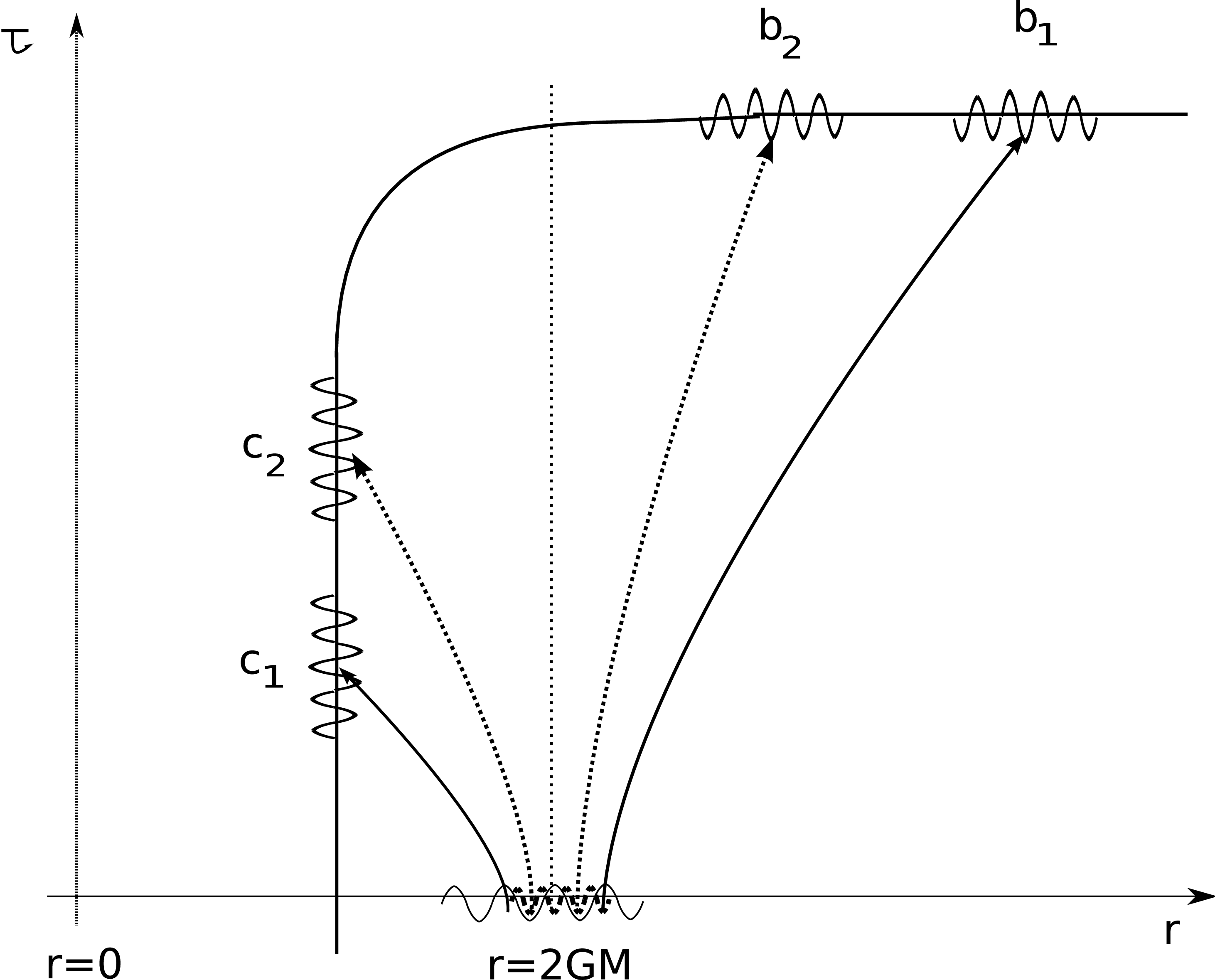}
%
%
\caption{On the initial spacelike slice we have depicted two fourier modes: the longer wavelength mode is drawn with a solid line and the shorter wavelength mode is drawn with a dotted line. The mode with longer wavelength distorts to a nonuniform shape first, and creates an entangled pairs $b_1, c_1$. The mode with shorter wavelength evolves for some more time before suffering the same distortion, and then it creates  entangled pairs $b_2, c_2$.}
\label{matftthree}       
\end{figure}

(2)  In a region with `solar system physics' we can identify and follow the evolution of an outgoing normal mode with wavelength $\lambda={M\over \mu}$ with $\mu>1$ a number of order unity (we depict this evolution of modes in figures \ref{matffourt},\ref{matftthree}). For concreteness, take $\mu=100$. Again using the fact that we are in the domain of standard solar system physics, we know that the state in this mode can be expanded in terms of  a Fock basis of particles. Thus when   $\lambda\sim {M\over 100}$ we can write
\be
|\psi\rangle_{mode}=\alpha_0|0\rangle+\alpha_1|1\rangle+\alpha_2|2\rangle\dots
\ee
There are two possibilities:

\b

(a) 
\be
\sum_{i>0}|\alpha_i|^2\sim 1
\ee
This means that there are particles with wavelength ${M\over 100}$ at the horizon. This means that the state at the horizon is not the vacuum, and so we do not have the traditional black hole, and are thus in violation of condition (ii) of the theorem. (Recall from the discussion in section (\ref{trhole}) that there is no ambiguity  in deciding if modes with wavelength ${M\over 100}$ are populated by quanta; this ambiguity comes only when $\lambda\sim M$.)
\b

(b) 
\be
\sum_{i>0}|\alpha_i|^2<\epsilon',~~ ~~~\epsilon'\ll 1
\label{bound}
\ee
In this case the state in the mode is the vacuum when $\lambda={M\over 100}$. The requirement of solar system physics tells us that the evolution of this vacuum mode will have to be agree with the leading order evolution of vacuum modes on this geometry to within some accuracy governed by a small parameter $\epsilon$. Thus there will exist an $\epsilon\ll 1$ such that  (\ref{cond1}) is satisfied by the evolution where the wavelength grows from $\lambda={M\over 100}$ to $\lambda\sim M$ and particle pairs populate this mode.

\b

(3) Since we have the niceness conditions N, the requirement of `solar system physics' under these conditions forces us to the fact that the particle pairs in option (b) above will be produced in a state close to the state $S^{(1)}$ in (\ref{set}). By Theorem 1, the entropy of entanglenment {\it increases} by at least $\ln 2-2\epsilon$ with each timestep. It is crucial that unlike the case of normal hot bodies, this entanglement entropy cannot start decreasing after the halfway evaporation point; this difference was discussed in detail at the end of section (\ref{deformation}).

\b

(4) The evaporation process produces $N\sim ({M\over m_{pl}})^2$ pairs before the hole reaches a size $\sim l_{pl}$. At this point we have a large entanglement entropy, for which we can write 
\be
S_{ent}>{N\over 2}\ln 2
\ee
since $\epsilon \ll 1$. Following the argument in section (\ref{haw}) we find that we are forced to mixed states/remnants (i.e. if the planck sized hole evaporates away we get a radiation state `entangled with nothing' violating quantum unitarity, and if a planck sized remnant remains, then we have to admit remnants with arbitrarily high degeneracy in the theory). 

\b

This establishes the Hawking theorem, \quad $\square$


\b

We have taken care to state Hawking's argument in a way that is a `theorem', so that if we wish to bypass the conclusion that we get mixed states/remnants then we have to violate one of the assumptions stated in the theorem. Thus we can either argue that the niceness conditions N need to be supplemented by further conditions (in which case we have to say what they are), or we have to argue that we do not obtain the traditional black hole in the theory (i.e. there will not be an information free horizon). 

\b

We emphasize the essential strength of Hawking's argument in the following corollary:
\b

{\bf Corollary 1}\quad {\it If the state of Hawking radiation has to be a pure state with no entanglement
with the rest of the hole then the evolution of low energy modes at the horizon has to be altered {\bf by order unity}. }

\b

The proof follows from Theorem 1. A small change in the state at the horizon changes this entanglement by only a small fraction, and cannot reduce it to zero. Conversely, if we wish this entanglement to be zero then we have to change the state of the created pairs to a state that is close to being {\it orthogonal} to the semiclassically expected one. \quad $\square$

\b

When people first come across Hawking's argument, it appears that there are several ways around it. So we now discuss aspects of the argument in more detail to understand why the information problem stayed unresolved for so long.

\section{Consequences of the Hawking `theorem'}

Given the theorem, we have three choices:

\b

{\bf 1. New physics:} We can accept the assumptions made in the theorem as natural, and thus accept the conclusion that some new physics will be encountered when black holes evaporate. 

\b

{\bf 2. No traditional holes:}  We can try to argue that the traditional hole does not form; i.e., the horizon has distortions (hair) which depend on the state of the hole. There were several attempts to look for such hair, by solving the linear wave equations for scalars, vectors, tensors etc. in the black hole background and looking for solutions with nontrivial support around $r=2M$. In general one finds that any such hair would have a divergent energy-momentum tensor at the horizon, and so would not be an acceptable solution. While the failure of these attempts did not prove that hair could not exist, the negative results came to be collectively dubbed the `no-hair theorem' for black holes, which says that the black hole geometry is characterized only by its global conserved quantum numbers like mass, charge and angular momentum.

\b

{\bf 3. Incompleteness of niceness conditions N:} We can argue that the niceness conditions N are insufficient to guarantee local Hamiltonian evolution, and must be supplemented with a new condition. The reason to do this would be the following. Consider a spacelike slice in fig.\ref{ftwo}. The matter $|\psi\rangle_M$ is $\sim 10^{77}$ light years from the place where the pairs are being created, but this  distance is one measured {\it along} the slice. We know that in some way this entire length of the slice is packed into the a  region $r\lesssim 2M\sim 3~{\rm Km}$, which is the radius of the hole as seen from outside. This is of course a peculiarity of the black hole geometry: because space and time interchange roles inside the horizon, we can make arbitrarily long spacelike slices $r=constant$ while staying inside the hole. We may therefore like to argue that the niceness conditions N do not guarantee local Hamiltonian evolution in gravity: we can have nonperturbative effects involving the entire region $r<2M$ which will `nonlocally' connect physics at one point of the slice with another point of the slice.

A gravitation instanton describing the black hole has been known from early days of black hole physics: the Euclidean `cigar' solution which has action
$
S_E=GM^2
\label{el}
$.
This suggests instanton effects of order
\be
 e^{-GM^2}\sim e^{-({M\over m_p})^2}
 \label{action}
\ee
But we need corrections {\it of order unity} in the state of entangled pairs. 
The traditional relativist can look at the state (\ref{qtwoq3}) obtained in his `nice slices evolution', and ask how instanton effects, continued back to Lorentzian signature, change this state to a state with no entanglement between the radiation and the hole. Until he is shown how such a change can happen, he is stuck with the Hawking theorem which gives mixed states/remnants at the end of evaporation.

\subsection{Mixed states and information}

In all our discussion of the Hawking theorem we have not used the term `information'. Even though the Hawking argument is usually called the `information paradox', the problem raised by his computation is not really centered on information, but rather on the mixed nature of the radiation state. In fact one can make radiation states that have full information about the hole but are still mixed  (and so violate unitarity), and conversely one can have the radiation state a pure unmixed state and yet carry no information about the hole. Let us see this in some detail since it leads to one of the main confusions about the nature of Hawking's paradox.

\b

{\bf Example 1:}\quad  Suppose the matter state is $|\psi\rangle_M=(\alpha|\uparrow\rangle+\beta |\downarrow\rangle)_M$. Suppose that the process of evolution creates {\it two} $b, c$ pairs, with the full state being described as follows:
\bea
\Big(\alpha|\uparrow\rangle+\beta |\downarrow\rangle\Big)_M&\r &\Big(\sq\u_M\d_{c_1}+\sq\d_M\u_{c_1}\Big )\otimes \Big (\alpha\u_{b_1}+\beta\d_{b_1}\Big )\cr
&&~~~~~~~~~~~~~~~~~~~~~~~~~\otimes\Big ( \sq \u_{c_2}\d_{b_2}+\sq \d_{c_2}\u_{b_2}\Big)
\eea
Note that we have made a toy model with a hypothetical evolution; the state on the RHS is nowhere near the state we get from semiclassical evolution. With this evolution, the first Hawking quantum $b_1$ carries the full information about the initial state, so the information comes `out'. But there is a second quantum $b_2$ which is entangled with $c_2$, so that the Hawking radiation has entanglement entropy $\ln 2$ with the matter inside the hole. If the hole evaporates away then the final state of radiation will be a mixed state, and we will get loss of unitarity even though the information has been retrieved. 

This toy example may appear unnatural because we let a  second Hawking quantum be emitted after the first one carried out all the information. But in fact the entropy of Hawking radiation quanta is some $\sim 30\%$ larger than the Bekenstein entropy of the hole \cite{zurek}, due to the fact that the radiation free-streams out of the hole instead of emerging quasi-statically in a reversible manner. Thus the number of emitted quanta is larger than the minimum number needed to carry the information of the hole.  

\b

{\bf Example 2:} \quad Let the initial matter state $|\psi\rangle_M=(\alpha|\uparrow\rangle+\beta |\downarrow\rangle)_M$ evolve as
\be
\Big(\alpha|\uparrow\rangle+\beta |\downarrow\rangle\Big)_M\r \Big ( \alpha \u_M\d_c+\beta \d_M\u_c\Big ) \otimes \Big( \sq \u_b+\sq \d_b\Big)
\ee
This time the state outside (given by the $b$ quantum) is a pure state with no entanglement with the state inside the hole. But this state carries no information about the initial matter state $|\psi\rangle _M$, so if the black hole disappears we will be left with a pure state and yet lose information.

\b

When we burn a piece of coal we have normal quantum evolution, so the radiation is in an unmixed state and also has the information of the coal. In black hole evaporation the state (\ref{qtwoq3}) has {\it both} the problems of examples 1,2: the radiation state is entangled with the state in the black hole interior, and also the radiation has only an infinitesimal amount of information about about the matter $|\psi\rangle_M$ (this infinitesimal amount of information arises from the small corrections of order $\epsilon$ that we have allowed). It is natural to expect that a solution to Hawking's paradox will resolve both problems at the same time, but it is helpful to keep in mind the above two examples when discussing information because the terms information loss and mixed state formation are sometimes used without distinction.

\section{AdS/CFT and the information paradox}

AdS/CFT \cite{maldacena, gkp, witten} duality is arguably one of the most interesting insights to emerge from string theory. It is also a very useful tool in understanding black hole behavior. But we cannot simply invoke this duality to bypass the information paradox. Since this is a very common confusion among students of string theory, we present it as the following discussion:
\b

Student: I dont see why I should worry about Hawking's paradox. Now that we know that gravity is dual to a CFT, and the CFT is unitary, there can cannot be any information loss, and so there is no problem.
\b

\hb That is an entirely circular argument, as I can easily show. Suppose I say: Quantum mechanics is unitary, so there can be no information loss. Would I have resolved Hawking's paradox?
\b
Student: No, that would be silly. Hawking agrees that quantum mechanics is valid in all laboratory situations. All he argues is that once we make a black hole, then quantum mechanics is violated. So we cannot use our tests of quantum mechanics in the everyday world to argue that there will be no problem when black holes form.
\b

\hb Good, that is correct. So now me let me ask the same question about AdS/CFT. You have computed the spectrum , 2-point functions, 3 point functions etc. and found agreement between the CFT and gravity descriptions. I understand that you have numerous such computations. But these processes do not involve black hole formation, and so do not address Hawking's argument. Is that correct?
\b
Student: Yes, that is correct. But we also have a black hole solution, called AdS-Schwarzschild, which is similar to the metric (\ref{ten}) in its essential respects.
\b
\hb Excellent. So I will now apply the Hawking theorem, proved in the above sections, and {\it prove} that normal assumptions about locality gives mixed states/remnants. Since your black hole has an `information free horizon' just like the Schwarzschild hole, my arguments go through in exactly the same way. Thus you have three choices:  (a) You can tell me why local Hamiltonian evolution breaks down under the niceness conditions N (b) you can  agree to mixed states arising from pure states, which violates quantum theory; in that case you lose AdS/CFT and string theory as well, since these are built on a foundation of usual quantum theory (c) You can agree to have remnants in your theory, and explain why they do not cause the problems that people feared. Now which will it be?
\b
Student: I don't know ... I see that you have forced me into a corner by using the Hawking theorem, and I will have to work as hard to solve it in my AdS case as I would have had to in the usual asymptotically flat case. So let me try to evade the problem by trying a different argument. I will use the CFT to {\it define} my gravity theory. Then I will get a gravity theory that has the expected weak field behavior, and I will never violate quantum mechanics, and I can never get information loss.
\b
\hb Excellent. With this definition of your gravity theory, you will by construction never have the `mixed state' possibility in Hawking's theorem. So now tell me: (a) Will you claim that traditional black holes do not form in this gravity theory (b) The black hole horizon forms, but the niceness conditions N do not give locality; in this case you should be sure to tell me how this happens and what niceness conditions you will add to recover conditions for the solar system limit (c) Do neither of the above but say that the theory has long lived remnants.
\b
Student: Well ... I always assumed that I could have a normal black hole horizon, usual notions of niceness conditions, and still get all the information out in Hawking radiation so there are no remnants. But I see now that the Hawking theorem forbids exactly this possibility. I dont know how I can say anything about the options you list without studying the black hole formation/evaporation process in detail in either the CFT or the gravity theory.
\b
\hb Exactly. You are welcome to do your analysis in either the CFT or the gravity theory, but at the end you must show me what happens when a black hole forms and evaporates in the gravity description. 
\b
Student: I see now that to solve Hawking's paradox I will have to understand the interior structure of the black hole. I cannot get by with any abstract argumments like `AdS/CFT removes the paradox.'
\b
\hb Exactly; in fact  abstract arguments in general cannot distinguish between whether locality broke down and information came out in the Hawking radiation or if information leaked out from a long lived remnant. Solving the information paradox implies that you tell us which happens, and if you want the information to come out in the radiation, to show explicitly the process by which `solar system physics' broke down while the niceness conditions N were still valid.
\b
\section{Resolving the paradox: results from string theory}

In recent years string theory has provided a set of results which together give a comprehensive picture of how information can come out of black holes. Let us list the results first, and then discuss  the qualitative picture of black hole formation and evaporation suggested by these results. 

\b

(1) {\bf Fractionation:}\quad In the early days of studying black holes in string theory it was assumed that at weak coupling we have a collection of branes, while at strong coupling we would get a black hole. This black hole was supposed to arise because the brane bound state stayed small (planck size of string size) while the gravitational radius became big, generating the traditional horizon around a central mass. But a closer analysis of string bound states shows that this is not the case, because of an effect termed `fractionation'. If we make a bound state of different kinds of branes, then the effective excitations occur in fractional units \cite{dasmathurpre}; these are therefore light, and if we estimate the `effective size' of the brane bound state then it grows with the number of branes as well as with the coupling \cite{emission}. This growth is such that the size of the brane bound state (for the D1D5P system \cite{sv}) remains order horizon size at all couplings where a black hole can form. This suggests a picture where we do {\it not} get a traditional black hole because the degrees of freedom of the hole distribute themselves throughout a horizon sized ball.

\b

(2) {\bf 2-charge microstates:}\quad The above picture was just based on a very crude estimate, but we can now start with simple holes and start constructing all their microstates. The simplest holes are extremal 2-charge holes: the D1D5 system. Entropy counts in string theory are based on using arguments that give a count in a weak coupling model, and using invariance of a supersymmetric index to argue that we have the correct count at strong coupling where a black hole is supposed to form. To solve the information paradox we need to understand the {\it structure} of the D1D5 system at at {\it strong} coupling where the black hole is supposed to form. 

How can we hope to do this? First, string dualities allow us to map the D1D5 system to the NS1P system: this is just a multiwound string (NS1) carrying momentum (P). Next, we start by looking at very simple states of this momentum carrying string. The momentum excitations are distributed like a 1-dimensional gas of excitations on the string. What if we take all the momentum to be in the lowest harmonic? This is certainly not a generic state, but it is a convenient starting point. It is analogous to studying the problem of electromagnetic radiation in a box, where the photons can be distributed among all normal modes. If we take a state where all the photons are in the lowest mode we get a `laser beam' which we can describe by classical $\vec E, \vec B$ fields, even though the generic state of the radiation cannot be so described since the fluctuation in these fields will be $O(1)$. For the string carrying momentum a similar thing happens, and the solution with all excitations in the lowest harmonic  is well described by supergravity fields even at strong coupling. We can write down the solution and transform it back by dualities to the D1D5 frame. This particular state of the D1D5 system is known to have a geometry {\it with no horizon and no singularity} \cite{bal}.

We can now move towards more generic states: for example we can distribute the momentum among {\it two} different harmonics. In the electromagnetic radiation case we still get well defined $\vec E, \vec B$ fields, though the fluctuations are slightly higher than before since the occupation number in each mode is somewhat less. For our problem of the string carrying momentum, the solution is still well described by a supergravity approximation (though the fluctuations will be slightly higher); again the solution (mapped back to the D1D5 duality frame)  can be explicitly written down and is found to have no horizon and no singularity \cite{lm4,lmm,internal}. 
An alternative description of these states can be obtained through the study of `supertubes'. 

This is interesting: when we look at these (nongeneric) states of the 2-charge D1D5 black hole we do not find the traditional black hole structure; instead the structure is modified all the way to where the horizon would have been, and in fact no horizon exists. Let us continue towards states that are still less generic: all we have to do is to continue distributing the momentum of the string among more and more modes till we reach the generic states described by bose and fermi distribution functions for the bosonic and fermionic excitations respectively. As we approach these generic states, we still see no formation of a horizon or singularity, but the {\it fluctuations} of the supergravity fields approach order unity. At the same time stringy corrections start appearing in the solution. First order corrections were analyzed in \cite{higher,phase} and interestingly, were found to be {\it bounded}; this is significant because if a correction diverged somewhere then it could convert a smooth solution to one with a horizon or a singularity. As it is, we just get a picture of a messy, quantum `fuzzball' as we approach the generic solution starting from the nongeneric ones. One now finds an interesting observation:  the surface area $A$ of this fuzzball satisfies a Bekenstein type relation \cite{lm5}
\be
{A\over G}\sim  \sqrt{n_1n_5}\sim S
\ee
where $S$ is the Bekenstein-Wald entropy of the  extremal hole made with $n_1$ D1 branes and $n_5$ D5 branes.

\b

(3) {\bf 3-charge microstates:}\quad The 2-charge hole is called the `small black hole', and one might wonder if similar constructions can be extended to 3-charge and 4-charge holes, which have `classical sized horizons'. Starting again from the nongeneric solutions and working towards less generic ones, we find a similar picture to the 2-charge case \cite{3charge, 3chargeother}. We write down all the microstates of the 3-charge hole at weak coupling, and start by finding the gravity description of the simplest ones at strong coupling.  Again, instead of getting a black hole with horizon, we get a solution with no horizon and no singularity. Large classes of such solutions are known now, and it is possible that there are enough to account for the entropy $\sqrt{n_1n_2n_3}$ of the 3-charge Strominger-Vafa hole \cite{bena}.

\b

(4) {\bf Hawking radiation:}\quad The information paradox began with the study of Hawking radiation. The above results on extremal holes show that the in string theory the microstates of the hole do not have the traditional horizon structure expected from the classical geometry. But extremal holes do not radiate, and it would be more satisfactory to explicitly see  the entire process by which black holes emit information carrying radiation. Following the same approach as above, we start with the simplest nonextremal microstate, and construct its geometry at strong coupling \cite{ross}. It is found that this geometry has an ergoregion, and thus emits energy by {\it ergoregion emission} \cite{myers}.   In \cite{chowdhurymathur} it was shown that this emission is exactly the Hawking radiation expected from this specific microstate. This agreement is very important, so let  us examine it in a little more detail. 

The D1D5 system is described by an `effective string', and the nonextremal D1D5 states carry left and right moving excitations along this effective string. Collision of these excitations lead to emission of radiation, with a rate which has the following schematic form:
\be
\Gamma(\omega)=V(\omega)\rho_L(\omega)\rho_R(\omega)
\label{gamma}
\ee
Here $V$ is the emission vertex, $\rho_L$ is the occupation number of the left moving excitations and $\rho_R$ is the occupation number of the right moving excitations. Each microstate has somewhat different $\rho_L, \rho_R$ in general, and will radiate a little differently. If we set $\rho_L=\rho^B, \rho_R=\rho^B$ where $\rho^B$ is the bose distribution function describing the generic excitation state for the given total energy, then it is known that $\Gamma$ gives the semiclassically expected Hawking radiation rate from the near extremal hole \cite{radiationall}. But this semiclassical computation of course gives a radiation that violates unitarity; its only the {\it rate} of emission $\Gamma(\omega)$ which agrees between the microscopic computation (which is unitary) and the gravity computation (which is not). But now we are in a position go do much more: we have an explicit construction of a nonextremal microstate at strong coupling (which is seen to have no horizon) and have computed its emission $\Gamma_{grav}$ in the gravity description (which happens by ergoregion emission). The microscopic description of this  microstate has particular values of $\rho_L=\bar \rho_L, \rho_R=\bar \rho_R$ (which are very nongeneric since this microstate is very nongeneric). We put $\bar\rho_L, \bar \rho_R$ in (\ref{gamma}) and compute the Hawking radiation $\Gamma_{micro}$ expected for this particular microstate. We find \cite{chowdhurymathur}
\be
\Gamma_{micro}=\Gamma_{grav}
\ee
To reiterate: {\it we  explicitly see the Hawking emission from a (nongeneric) microstate of the extremal hole; we can look at this radiation  and see how it carries out the information of the state.} This imprinting of information is straightforward in this simple example: the microstate has rotation, and the spectrum of radiation depends on the size and orientation of the ergoregion inside the geometry.

\b

{\bf Lessons from the dual CFT:} \quad We should also ask what we can learn from the dual CFT about dynamical  black hole formation and evaporation processes. In \cite{compmathur} it was found that if one uses the CFT at its free orbifold point, then an interesting thing happens as we track the evolution of a particle till it crosses the position where the horizon would have been naively expected: the notion of {\it where} the particle is becomes ill defined. If we use a definition of time which is not the one at infinity, but adapted to the `infalling frame', then we see no change as we cross the location where the horizon would have been, but such a time coordinate ceases to be definable after the infall time to the center of the hole. This suggests connections to the idea of `complementarity', something which should be explored further.

If we wish to study not a single particle infall but the collapse of an entire shell to make a black hole, then we have to do more: the free orbifold CFT does not have the interactions needed to thermalize the 
initial state to a generic state of the hole. Thus we have to use the `deformation operator' which take the CFT away from the orbifold point and makes it interacting. This operator twists together two strands of the string, adding an insertion of the supersymmetry operator at the vertex. Qualitatively, we can see that such an operator will take a pair of quanta in the CFT, and upon their collision, distribute their energy over several quanta. This process will therefore take a pair of high energy excitations (which would be created by a single high energy particle falling into the AdS region of the dual geometry) and convert them to lower and lower energy quanta, a process which looks like the thermalization of the initial quantum into a superposition of fuzzballs. One can ask when in the gravity theory there is enough energy to create a black hole; it was found in \cite{lm4} that this energy is such that if we distribute it into lower energy quanta in the CFT then these CFT excitations would cover the entire `effective string'.  Thus we get a simple description of black hole formation in the dual CFT: the initial excitations created by an infalling particle break up into lower energy excitations; if this process ends before the entire effective string is covered, we do not get a black hole. (The process can end before black hole formation because we have a given length for the effective strings in the initial state, and a string of a given length $L$ cannot support quanta of energy lower  than ${2\pi\over L}$.) One notes in this computation that the criterion for black hole formation depends only on the length of the components of the effective string. and not on the effective coupling. Thus this thermlization process is not `stringy', rather it involves multiple insertions of the gravitational interaction vertex in the gravity description. This suggests that the formation of fuzzballs happens when the large phase phase space of fuzzball solutions is explored by gravity interactions, and not by $\alpha'$ corrections to the metric.

It may seem that such a computation in the CFT using  the deformation operator will be very complicated since the entire process of black hole formation will involve a large number of interactions. What helps here is that as we increase the coupling, the rate at which thermalization can take place does not keep growing: it saturates because if we wish to create quanta of energy $\omega/2$ from quanta of energy $\omega$ then the process will take a time at least $O(1/\omega)$ because of phase space constraints. Since the CFT has no intrinsic length or time scale,  the time for the mean energy $\omega$ of quanta to drop to say half its value must be determined by $\omega$ alone (if it is going to remain nonzero as the effective coupling $g_{eff}\r\infty$). Thus we get a simple picture of the thermalization process: quanta split into other quanta as fast as they can given their wavelengths, and the energy keeps getting redistributed over more and more strands of the effective string. The interesting thing here is that {\it any} strand of the effective string can interact with {\it any} other strand, so we have a huge (order $n_1n_5$) coordination number for the interaction, unlike normal physics where a particle can only talk to a few nearest neighbours. This picture agrees with the requirement in \cite{hayden} that the thermalization be exponentially fast to save the idea of complementarity; this is something that should be explored in more detail.

\b

{\bf Phase space:} \quad Finally, let us come to the intuitive picture of why black hole microstates can be quantum fuzzballs when the niceness conditions N required the traditional black hole with an `information-free horizon'. The crucial issue appears to be the analysis of {\it phase space}. In the 2-charge solutions made from a string carrying momentum, we find that an interesting fact of string theory was crucial: there are no longitudinal waves on the fundamental string. Thus if we try to add the second charge (momentum) to the first charge (the string), then we necessarily have to make the string execute transverse oscillations. This is what causes the string to spread over a nontrivial transverse area, and ultimately to fill up a horizon sized region. Why does the string need this much transverse space? Suppose we require the string to execute its oscillations in a much smaller transverse radius. Then the momentum will have to be carried by a few high frequency modes, and there is a only a small number of such states \cite{review}. In general we find that if we allow too small a transverse space for the states to live in, then we cannot fit in the full phase space of solutions contributing to the Bekenstein entropy \cite{phase}.
Phase space arguments also tell us that the infinite throat of the classical 3-charge solution will be cut off at some depth since the infinite throat is only one point in a space of `capped throats' \cite{phase}. A detailed analysis of phase space measures for the 3-charge case was done in \cite{deboer2}, and it was found that the throats will end at a depth expected by the microscopic analysis of states.

We now have all the  tools that we need to put together a logical picture of black hole formation and evaporation. Suppose a shell collapses to form a black hole. The time independent states at energy $M$ are fuzzballs, with no horizon. Let us estimate a tunneling amplitude between the shell state and any of the fuzzball states. These are both macroscopically different states, so the amplitude ${\cal A}$ for this tunneling is very small, as was noted in (\ref{action}):
\be
{\cal A}\sim  e^{-S_{tunnel}}, ~~~S_{tunnel}~\sim~ {1\over 16\pi G}\int R~\sim~ GM^2\sim ({M\over m_p})^2
 \label{actionq}
\ee
But now comes an interesting twist: the {\it number} of fuzzball states  ${\cal N}$ that we can tunnel to is very large, being given by the Bekenstein entropy
\be
{\cal N}\sim e^{S_{bek}}\sim e^{GM^2}
\ee
Thus we see that the smallness of the tunneling amplitude ${\cal A}$  can be offset by the largeness of ${\cal N}$ \cite{offset}. The large entropy is of course a characteristic feature of black holes, not present in other macroscopic bodies like the sun. In any path integral we have the action, and a measure. Usually the measure factor is `small' (i.e. order $\hbar$) in its effect on macroscopic processes, but the large entropy of the black hole has made its effects comparable to the effects of the classical action, and thus invalidated classical intuition. 

We need one last estimate before the picture is complete. Tunneling is really a spreading of the wavefunction over available phase space, and we need to know that this happens in a time shorter than the Hawking evaporation time; otherwise we would not have avoided Hawking's paradox and would end up with remnants. Interestingly, a simple estimate of this spreading time shows that it will always be less that the Hawking evaporation time \cite{release}.

To summarize, we have put together three things in this picture: (a) We have recognized that black hole microstates are fuzzballs with no traditional horizon (b) An estimate of the tunneling process shows that the smallness of the amplitude to tunnel to fuzzballs can be offset by the largeness of the number of states available to tunnel to (c) The time for this tunneling is shorter than Hawking evaporation time, so the collapsing shell becomes a linear combination of fuzzball states quickly enough so that radiation can now emerge from the fuzzball rather than a traditional horizon.

\section{Lessons from the information paradox}

The first point to note about the information paradox is that it is very nontrivial: we {\it cannot} preserve the standard rules for when we have nice local evolution, and still avoid Hawking's puzzle. It is a common misconception that small corrections to Hawking radiation, ignored in the leading order computation, will resolve the problem; we proved explicitly in Theorem 1 that such is {\it not} the case. The resolution of the puzzle is much more nontrivial and involves a deep lesson of quantum gravity: there can be such a large phase space of solutions in a given region that path integral is no longer dominated by the extremum of the classical action. Put another way, the error in earlier black hole work was to assume that quantum gravity effects are always confined to within the planck distance $l_p$. While $l_p$ is the natural length scale that we can make from $G , \hbar, c$, we need a large number of quanta $N$ to make the black hole, and we  have to then ask if quantum gravity effects stretch over distances $l_p$ or distances $N^\alpha l_p$ for some $\alpha>0$. One finds from the fuzzball computations that the latter is the case, and $\alpha$ is such that the effects reach upto horizon scales.

A common question is whether we can now conjecture a picture for the interior of the Schwarzschild hole. Let us draw our intuition from the nonextremal microstate
described in the last section. The crucial point is that the  geometry was independent of $t$, yet we cannot make a spacelike slicing that is time-independent. This is because we have an ergoreion; the killing vector $\p/\p t$ is timelike at infinity, {\it but is not timelike everywhere.} Thus the $t=constant$ surface is not spacelike everywhere.  If we do try to foliate the spacetime with spacelike slices then we find that these slices `stretch' during the evolution, and create particles leading to particle creation by the mechanism of ergoregion emission. This emission agreed {\it exactly} with the Hawking radiation expected for this microstate. We can now move towards more  generic microstates, just like we did for the extremal case: we get more and more complicated ergoregions \cite{cm2}. 
The general classical microstate is found to always have either an ergoregion or time dependence, and so radiates energy. As we reach the generic state the scale of variation of the features in the geometry is expected to reach the planck scale, and stringy effects will give us a quantum fuzzball, just like in the extremal case. 

Why doesn't the energy of the Schwarzschild hole  all fall into the origin at $r=0$? Consider the simpler case of a single string in flat space. It would seem that a string in the shape of a circle must shrink to a point under its tension so there should be no extended string states. But of course we can get    extended states: the string profile is not circularly symmetric, and while each segment of the string is indeed trying to shrink, the whole exited string maintains a nonzero size in its evolution. Similarly, the nonextremal microstates are not spherically symmetric, and they cannot be sliced in a time independent manner. Each part of the geometry is dynamical, and the whole structure maintains a nontrivial structure without generating a traditional horizon.

It would be interesting to apply these lessons to Cosmology. Near the big bang singularity, we have a very high density of matter, and thus presumable a large available phase space of states. Spreading over phase space may again make it incorrect to take any one of these states and evolve it classically. One approach to analyzing the early Universe using the full entropy of states was taken in \cite{cmuniverse}. It would be interesting to analyze possible relations between the large space of BKL singularities and the possible microstates at the initial singularity; there are strong similarities between the local behavior at the BKL singularity and the pole structure of fuzzball states. In general, it would indeed be very exciting to explore relations between  black holes and the physics of the early Universe. 

\section*{Acknowledgements}

 Not everyone understands Hawking's paradox the same way, and discussing the paradox with people having different views has helped me enormously in making this hopefully precise version of Hawking's argument. In particular I would like to thank Sumit Das, Gary Horowitz, Don Marolf, Don Page, Ashoke Sen, Steve Shenker, Lenny Susskind and Edward Witten for patiently debating many issues about the paradox. I thank Patrick Hayden for discussions and for suggesting the use of entropy inequalities. For the work of fuzzballs which I summarized at the end, I would like to thank my collaborators Steve Avery, Borun Chowdhury, Stefano Giusto, Oleg Lunin, Ashish Saxena and  Yogesh Srivastava, and all my colleagues who have worked tirelessly on this problem. Finally I am grateful to Angel Uranga and all the other organizers of the CERN school for giving me the opportunity of presenting these ideas there. 
 
 This work was supported in part by DOE grant DE-FG02-91ER-40690.

\end{document}